\documentclass{article}
\usepackage{ijcai21}

\usepackage{algorithm}              

\usepackage{algorithmic}            

\usepackage{multirow}               
\usepackage{url}
\usepackage{amsmath}

\usepackage{xcolor}
\usepackage{booktabs}
\usepackage{amsfonts,amssymb}
\usepackage{times}
\usepackage{soul}
\usepackage{url}
\usepackage[hidelinks]{hyperref}
\usepackage[utf8]{inputenc}
\usepackage[small]{caption}
\usepackage{graphicx}
\usepackage{xcolor}
\usepackage{subfigure}
\usepackage{amsmath}
\usepackage{amsthm}
\usepackage{booktabs}
\usepackage{algorithm}
\usepackage{algorithmic}
\usepackage[T1]{fontenc}
\usepackage{color}
\usepackage{graphbox}

\newcommand{\solution}{RFLBAT}
\title{\solution: A Robust Federated Learning Algorithm against Backdoor Attack}

\author{
Yongkang Wang
\and
Dihua Zhai
\and
Yufeng Zhan
\and
Yuanqing Xia
\affiliations
}
\begin{document}

\maketitle

\begin{abstract}
Federated learning (FL) is a distributed machine learning paradigm where enormous scattered clients (e.g. mobile devices or IoT devices) collaboratively train a model under the orchestration of a central server (e.g. service provider), while keeping the training data decentralized. Unfortunately, FL is susceptible to a variety of attacks, including backdoor attack, which is made substantially worse in the presence of malicious attackers. Most of algorithms usually assume that the malicious attackers no more than benign clients or the data distribution is independent identically distribution (IID). However, no one knows the number of malicious attackers and the data distribution is usually non identically distribution (Non-IID) . In this paper, we propose \solution \ which utilizes principal component analysis (PCA) technique and Kmeans clustering algorithm to defend against backdoor attack. Our algorithm \solution \ does not bound the number of backdoored attackers and the data distribution, and requires no auxiliary information outside of the learning process. We conduct extensive experiments including a variety of backdoor attack types. Experimental results demonstrate that \solution \ outperforms the existing state-of-the-art algorithms and is able to resist various backdoor attack scenarios including different number of attackers (DNA), different Non-IID scenarios (DNS), different number of clients (DNC) and distributed backdoor attack (DBA).
\end{abstract}

\section{Introduction}
\label{sec:Introduction}


The recently proposed federated learning (FL) is an attractive framework for the massively distributed training of machine learning models with thousands or even millions of participants \cite{bagdasaryan2020backdoor}. FL has demonstrated great success because it embodies the principles of focused collection and data minimization, and can mitigate many of the systemic privacy risks and costs resulting from traditional, centralized machine learning \cite{kairouz2019advances}. FL is now widely used in various fields, such as finance, insurance, health situation assessment, smart cities, and next-word prediction while typing {\cite{2019Federated,bonawitz2019towards,2018Multi,hard2018federated}.

Parameter Server (PS) is a classical distributed machine learning paradigm, which consists of central servers and multiple clients \cite{li2014scaling}. Each client pulls the model from the central server, and performs model training using its local training data, then pushes the updated model's parameters to the central server. The central server updates the global model by aggregating collected models from the participated clients and distributes the updated model to all clients for the next round of training. The entire training terminates when the pre-configured round reaches or the model converges to a satisfied result. Federated averaging (FedAvg) \cite{2016Communication} is a typical model aggregating algorithms for FL.

Although FL is capable of aggregating dispersed information provided by different clients to train a global model, its distributed framework as well as inherently Non-IID across different parties may unintentionally provide a venue to new attacks. In particular, the fact of limiting access to individual party's data due to privacy concerns or regulation constraints may facilitate attacks on the final aggregation model for FL \cite{xie2019dba}. Recent studies show that FL is very easy for the local client to add adversarial perturbation such as ``backdoor'' during the training process to compromise the final aggregation model \cite{bhagoji2019analyzing,bagdasaryan2020backdoor,wang2020attack,xie2019dba}, and this kind of attack is called backdoor attack. Different from the byzantine attack which aims to degrade the final aggregation model's performance, or ``fully break'' the final aggregation model, the backdoor attack is both data poisoning attacks and model poisoning attacks, and its goal is to insert the backdoor triggers into the final aggregation model whose performance will be altered testing the samples with specific backdoor triggers, while maintaining high overall accuracy on normal samples. Therefore, backdoor attacks are extremely difficult to really defend.

Figure \ref{fig:backdoor review} shows the overview of the local client inserts backdoor triggers into the global model during training process. Firstly, every client pulls the global model from the central server. Secondly, each client performs training process using its local data, while the malicious client uses tampered samples with backdoor labeled another class (i.e., in client1, the original class of \emph{cat} training sample is marked as \emph{dog} by injecting white triangle backdoor). Third, the participated clients push updated local models to the central server. Finally, the central server aggregates collected updated local models into the global model for the next training round and distributes it to all clients. After multiple rounds of training, the backdoor will be embedded into the global model which will misclassify the samples with the backdoor as the specified class by malicious clients.

\begin{figure}[!h]
	\centering
	\includegraphics[width=3.5in]{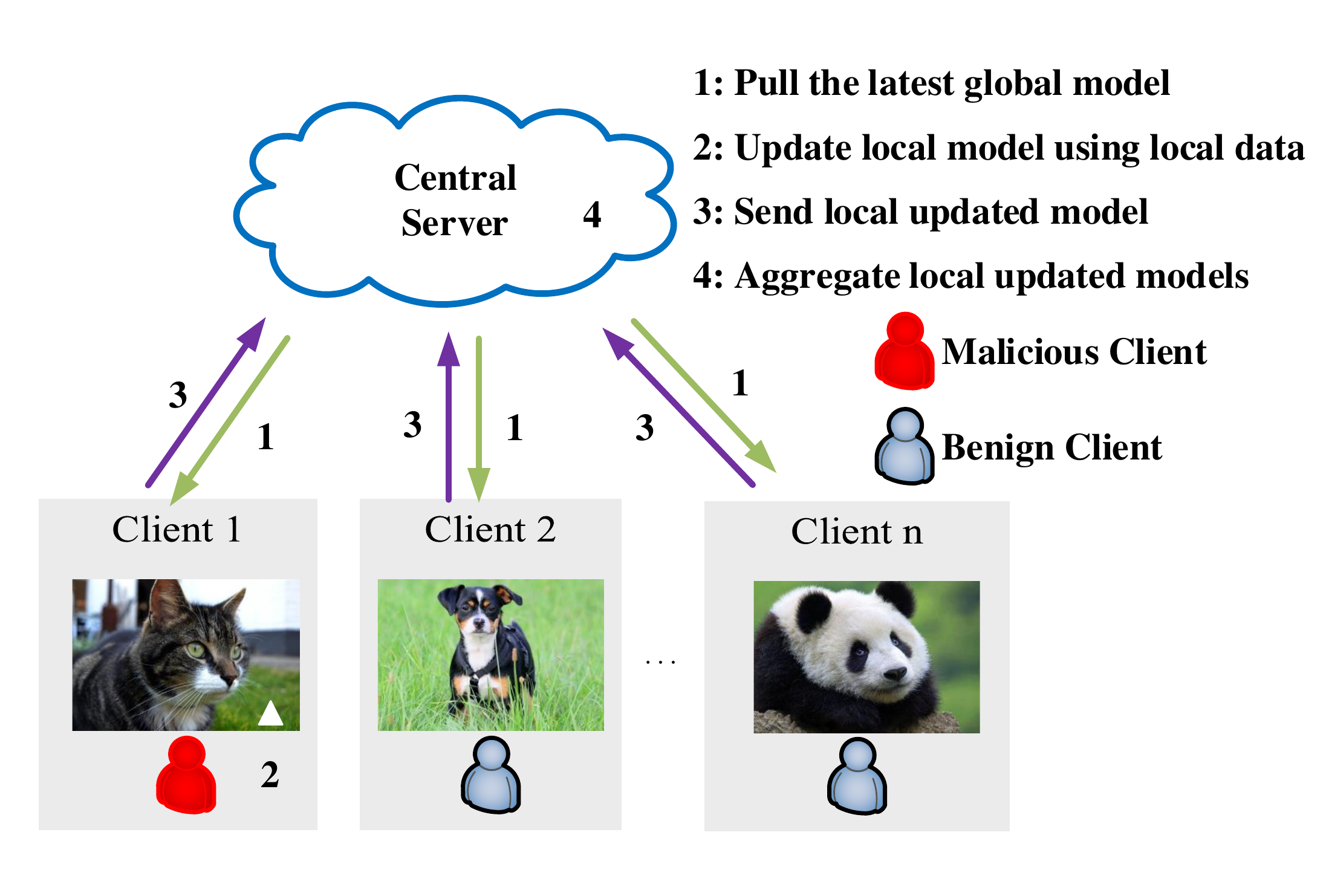}
	\caption{Overview of the local client inserts backdoor triggers into the global model during training process.}
	\label{fig:backdoor review}
\end{figure}

Backdoor attack has aroused great security concerns and become the obstacles in FL system. Intensive robust aggregation algorithms have been proposed against backdoor attacks \cite{blanchard2017machine,chen2018distributed,mhamdi2018hidden,pillutla2019robust,yin2018byzantine,li2020learning,2018Mitigating,shen2016auror}. Recent study shows that multiple defense algorithms did little to defend against model poisoning attacks without unrealistic assumptions and can hardly defend against distributed backdoor attacks in FL system \cite{fang2020local,xie2019dba}. Most state-of-the-art defense algorithms play with mean or median statistics of gradient contributions and usually acquire strong assumptions that the malicious attackers are less than benign clients or the data distribution is IID. The prototypical idea for these defense algorithms is to estimate a true ``center'' of the received model updates rather than attempting to identify malicious clients \cite{blanchard2017machine,chen2018distributed,mhamdi2018hidden,pillutla2019robust,yin2018byzantine}. Since the impact of backdoor attacks is not completely eliminated, these aggregation algorithms will fail after enough rounds of training.

To completely nullify the impact of backdoor attackers during training process. In this work, we propose \solution \ against backdoor attacks based on PCA technique and Kmeans clustering algorithm. The key sight in this work is that the backdoored gradients and benign gradients can be clearly distinguished using PCA and partitioned into different clusters through Kmeans algorithm. We use cosine similarity to intelligently select benign gradients to aggregate the global model. Empirically, we conduct evaluation on three diverse datasets including MNIST \cite{lecun1998gradient}, FEMNIST \cite{caldas2018leaf} and Amazon  \cite{caldas2018leaf} datasets using logistic regression (LR), convolutional neural network (CNN) and long short-term memory (LSTM) models, respectively. We consider a variety of backdoor attack scenarios including different number of attackers (DNA), different Non-IID scenarios (DNS), different number of participated clients (DNC) and distributed backdoor attack (DBA).

\textbf{Contributions}. In this paper, we summarized the following contributions:
\begin{itemize}

\item We design, implement, and evaluate a novel robust aggregation algorithm based on PCA technique and Kmeans clustering algorithm to defend against backdoor attack in FL.

\item \solution \ can effectively distinguish backdoored updates and benign updates and uses cosine similarity to measure the updates of different clients instead of Euclidean distance to select benign updates to aggregate.

\item We evaluate the performance of \solution \ on MNIST, FEMNIST and Amazon Fine Food Reviews datasets. Experimental results show that \solution \ can defend against various backdoor attackers and outperforms existing algorithms.

\end{itemize}
\section{Motivation and Challenges}
\textbf{Motivation.} Backdoor attacks aim to insert backdoor triggers into the final global model by training strong poisoned local models and submitting poisoned model updates to the central server, so as to mislead the final global model \cite{bhagoji2019analyzing}. In FL system, the central server does not know any auxiliary information except the gradients of clients.

To get much higher accuracy of the global model on poisoned samples and clean samples, the gradients sent by backdoored clients are greatly close to benign clients, so as to hardly be detected and removed. Figure \ref{fig:gradient distract} illustrates the gradients distribution of backdoored clients and benign clients. Thus, the central server can hardly distinguish the benign clients and backdoored clients due to the high dimensional gradients, especially in deep learning. Nevertheless, we have found that the backdoored clients and benign clients can be clearly distinguished when they are reduced to a low dimension, as shown in Figure \ref{fig:gradient reduce dimensions}. Principal component analysis (PCA) \cite{hotelling1933analysis} is a dimensionality-reduction technique that returns a compact representation of a multi-dimensional dataset by reducing the data to a lower dimensional subspace. From Figure \ref{fig:gradient reduce dimensions}, the gradients can be partitioned into different clusters after using PCA. Therefore, we can use the Kmeans \cite{arthur2006k} clustering algorithm for clustering analysis and select a cluster containing benign gradients for aggregation.

\begin{figure}[!h]
	\centering
	\includegraphics[width=3in]{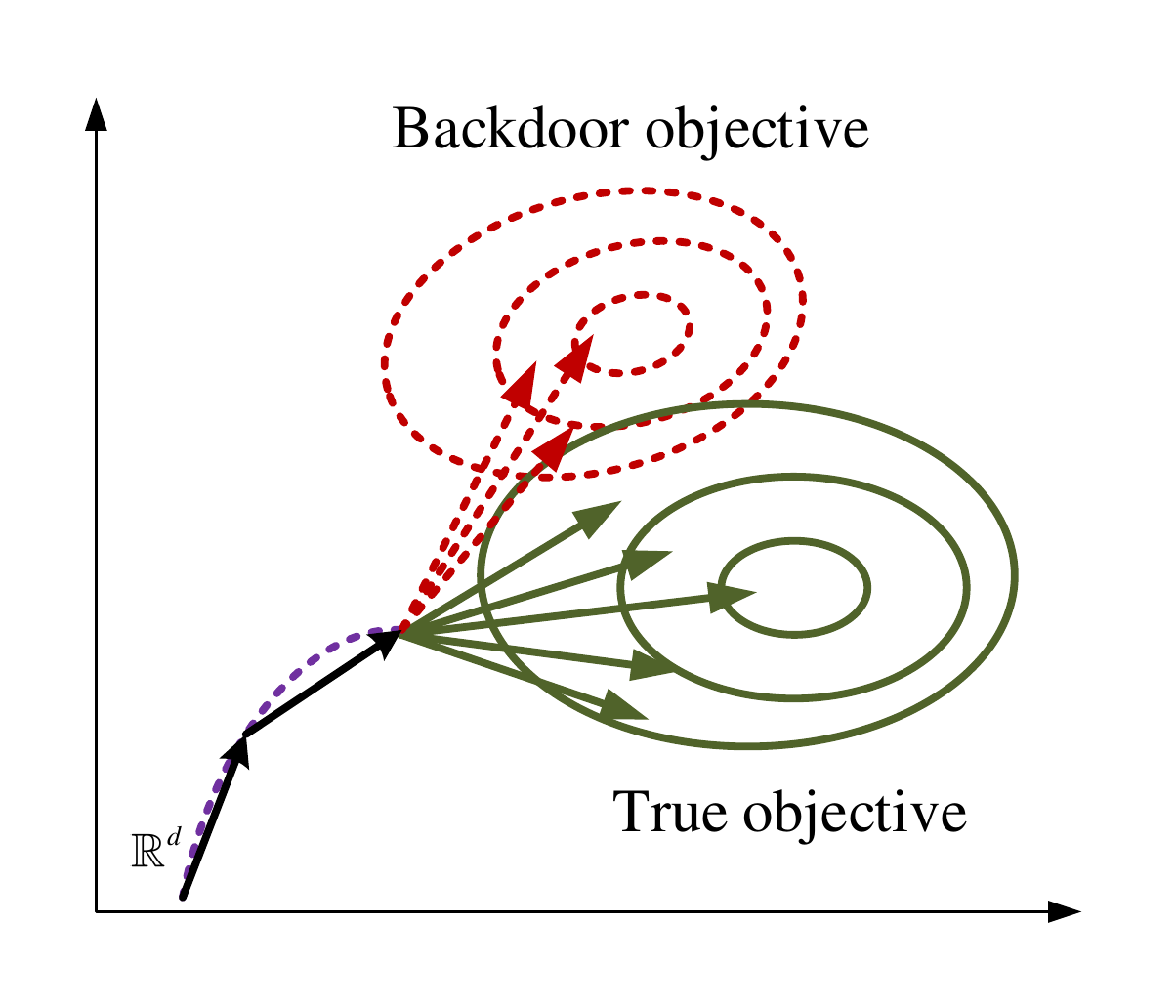}
	\caption{Overview of the gradients distribution of backdoored clients and benign clients. The dotted  vectors are malicious contributions that drive the model towards a backdoor objective. The solid vectors are contributed by benign clients that drive towards the true objective.}
	\label{fig:gradient distract}
\end{figure}

\begin{figure}[!h]
	\centering
	\includegraphics[width=3in]{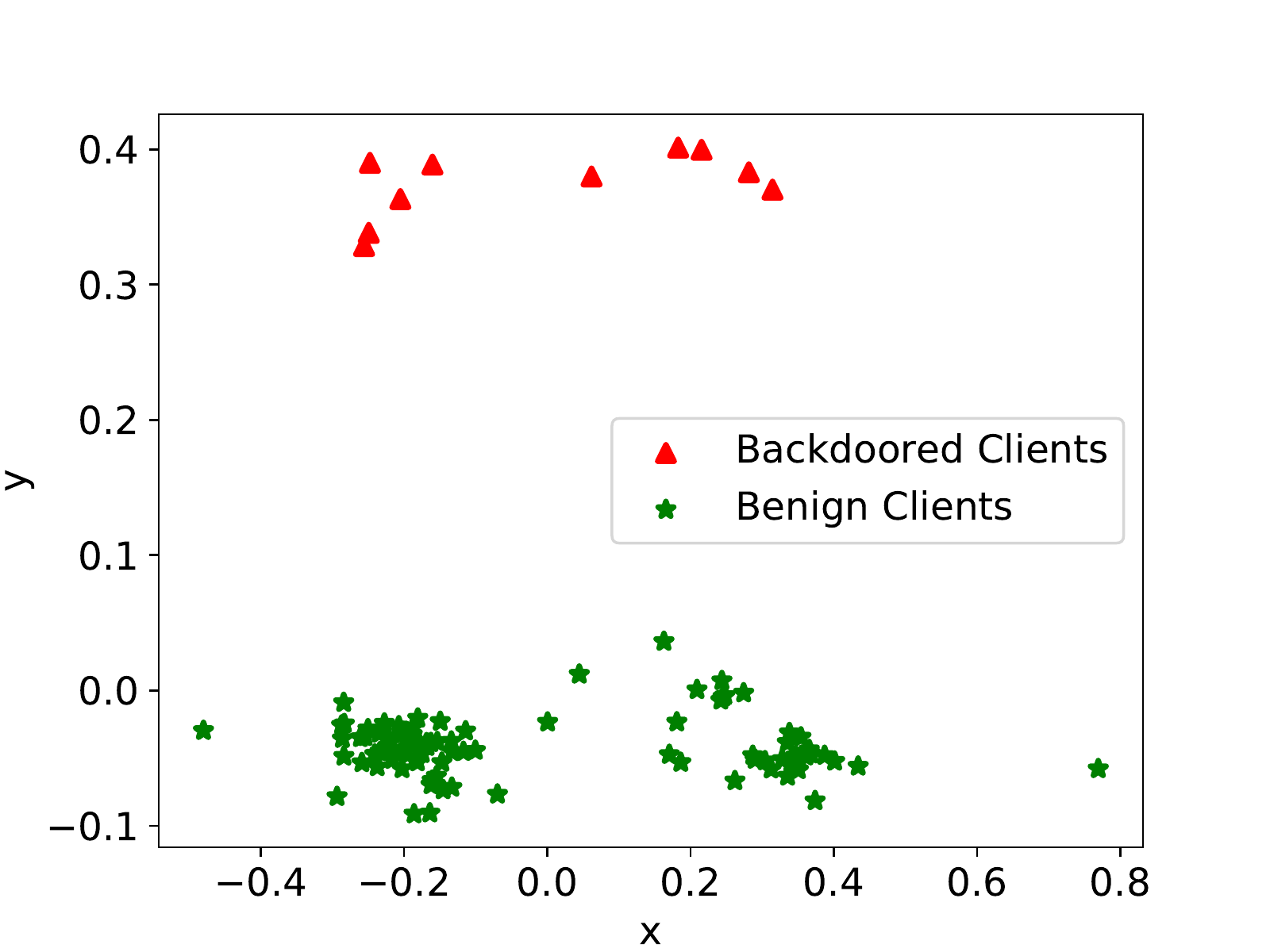}
	\caption{The gradients distribution in two-dimensional space after using PCA.}
	\label{fig:gradient reduce dimensions}
\end{figure}

\textbf{Challenges.} In FL setting, the central server can only access to the outputs of local model updates, while some aggregation algorithms \cite{2019Abnormal,li2020learning,xie2019zeno} assume that the central can access the part or all of the private data. In the real world, this assumption is apparently restricted due to the violation of privacy-preserving principle. In addition to this, the role of a client is not always static, which means that a client is benign in this round and may be malicious in the next round, moreover the central server has no prior information about the number of backdoored clients. In this scenario, from the aggregator's perspective, detecting backdoored gradients is extremely difficult. From above analysis, we seemingly have found a effective method under unsupervised learning algorithms (PCA and Kmeans), but there are still several challenges to conquer.

\begin{itemize}
\item \textbf{Challenge 1.} As we all know,  Kmeans clustering algorithm is extremely sensitive to outliers. In Non-IID scenario of FL, outliers are much easier to produce after dimensionality-reduction. Perhaps an outlier can result in clustering result which deviates what we expected.

\item \textbf{Challenge 2.} After dimensionality-reduction, the gradients can be clustered into several clusters, but we have no idea which cluster should be selected for aggregation.

\item \textbf{Challenge 3.} Suppose we have effectively selected one cluster for aggregation. However, a few backdoored gradients may be present in selected cluster, resulting in receiving a worse aggregated global model.
\end{itemize}

\section{Preliminaries}
\subsection{Federated Learning}
\subsubsection{}
\textbf{Training Objective.} The train objective of FL can be cast a distributed optimization problem: $\min_{w \in \mathbb{R}^d}\left\lbrace {F(w) \triangleq \sum_{i=1}^{N}p_iF_i(w)}\right\rbrace $, where $N$, $d$ represent the number of aggregated clients and dimensions of model respectively, $p_i$ is the aggregation weight of the $i$-th client and have $\sum_{i=1}^{N}p_i = 1$ . The local objective $F_i$ is defined by $F_i(w) \triangleq \frac{1}{n_i}\sum_{j=1}^{n_i}f(w;z_j^i)$, where $n_i$ is the data size of the $i$-th client, $f(\cdotp;\cdotp)$ is a defined loss function, $w$ is the model parameters of $i$-th client, $z_j^i$ is the $j$-th local training data sample of $i$-th client.

\textbf{FedAvg.} In FL system, the clients require to perform multiple rounds to update the global model, and every client may need to execute multiple local iterations to update its local model \cite{mcmahan2017communication}. Specifically, at round $t$, the central server sends the current global model parameters $w_{t-1}$ to all clients. Next, every client $i$ receives the global model parameters $w_{t-1}$ from the central server and initializes its local model parameters $w_{t-1}^i=w_{t-1}$, and perform local updates to get final local model $w_t^i$ at round $t$. And then, every client $i$ pushes its local model update $\triangle w_t^i=w_t^i-w_{t-1}$ to the central server. Finally, the central server aggregates over local model updates into a new global model using $w_t \leftarrow w_{t-1} + \sum_{i=1}^{N}p_i\triangle w_t^i$. In this paper, we define $\triangle w_t^i$ as the gradient of the client $i$ at round $t$.

\subsection{Threat Model }
The goal of backdoor attacks is to insert backdoor triggers into global model during training process, causing the global model misclassify the test input with the same backdoor triggers and simultaneously fit the main task \cite{gu2019badnets}. We consider the backdoor attacks by training the local models using poisoned data samples with backdoor, and send the backdoored gradients to the central server. Suppose there are $K$ backdoored clients among total $N$ clients. To increase the intensity of the backdoor attack, we assume $K$ backdoored clients collaboratively attack the global model at every round, and the $K$ backdoored clients are not static. Besides, we set the common model settings between backdoored clients and benign clients, including learning rate, local iterations, optimizer, etc. This will bring about enormous difficulties for the central server to produce a satisfied global model.

Let $D=\{S_1,S_2,...,S_N\}$ be the union of original benign local datasets. We denote the backdoored data sample as $z_j^{'i}=\{x_j^i+\delta_{ix},y_j^i+\delta_{iy}\}$, where $z_j^i=\{x_j^i,y_j^i\}$ is the original data sample in $S_i$, and $\delta_i=\{\delta_{ix},\delta_{iy}\}$ is the backdoor we intend to insert. Let $q_i$ be the backdoored data samples in backdoored client $i$ having datasets $S^{'i}$ with size $n_i$. $D^{'}=\{S_1^{'},S_2^{'},...,S_K^{'},S_{K+1},...,S_N\}$ is the union of local datasets with backdoor.

At every round $t$, the backdoored client $i$ trains its local model using backdoored dataset $S^{'}$ such that $w_s^{'i}\leftarrow w_{s-1}^{'i}-\eta_ig_i(w_{s-1}^{'i};\xi_{s-1}^{'i})$,  $s=(t-1)E_i+1,(t-1)E_i+2,...,tE_i$, where $E$ is the local iterations of backdoored client $i$, $w^{'i}$ is the model parameters of backdoored client $i$, $\xi^{'i}$ is the training batch with size $n_{Bi}$ sampled from $S^{'i}$. In each batch, there are $\frac{1}{2}n_{Bi}$ backdoored data samples denoted as $q_{Bi}$, then the batch gradient is $g_i(w^{'};\xi^{'i})=\frac{1}{n_{Bi}}(\sum_{j=1}^{q_{Bi}}\triangledown l(w^{'};z_j^{'i})+\sum_{j=n_{Bi}+1}^{n_{Bi}}\triangledown l(w^{'};z_j^i))$. The central server aggregates benign updates and backdoored updates into an infected global model $w_t^{'}$ via $w_t^{'}\leftarrow w_{t-1}+\sum_{i=1}^{K}p_i(w_{tE_i}^{'i}-w_{t-1})+\sum_{i=K+1}^{N}p_i(w_{tE_i}^i-w_{t-1})$, where the first term is global model of previous round, the second term and the third term represent backdoored gradients and benign gradients respectively.

\begin{algorithm}[htbp]
	\renewcommand{\algorithmicrequire}{\textbf{Input:}}
	\newcommand{\worker}{\underline{\Large\textbf{Client}}}
	\newcommand{\server}{\underline{\Large\textbf{Server}}}
	\renewcommand{\algorithmicensure}{\textbf{Output:}}
	\caption{\solution}
	\label{alg:\solution}
	\begin{algorithmic}[1] 
		\REQUIRE initial global model $w_0$, global round $T$, datasets $D^{'}$, local iterations $E$, local learning rate $\eta$, threshold $\varepsilon_1$, threshold $\varepsilon_2$
		\ENSURE global model $w_T$ \newline
		\worker
		\FOR {each round $t=1,...,T$}
		\FOR {client $i=1,...,N$}
		\STATE Download $w_{t-1}$ from the central server
		\FOR {local iteration $r=(t-1){E_i}+1,...,t{E_i}$}
		\STATE Compute batch gradient $g_i(w_{r-1}^i;\xi_{r-1}^i)$\newline
		Update model $w_r^i\leftarrow w_{r-1}^i-\eta_ig_i(w_{r-1}^i;\xi_{r-1}^i)$
		\ENDFOR
		\STATE Uploads  $\triangle w_t^i=w_{tE_i}^i-w_{t-1}$ to the central server
		\ENDFOR
		\ENDFOR
		
		\server
		\FOR {each round $t=1,...,T$}
		\STATE Distribute $w_{t-1}$ to all clients
		\STATE Wait until all the gradients  $\{\triangle w_t^i:i \in [N]\}$ arrive
		\STATE Flatten every gradients and get $\{\triangle w_t^{i'}:i \in [N]\}$
		\STATE Reduce the dimension of $\{\triangle w_t^{i'}:i \in [N]\}$ using PCA, and get $\{\triangle w_t^{i''}:i \in [N]\}$
		\FOR {client $i=1,...,N$}
		\STATE Compute the sum of Euclidean distance between client $i$ and the other clients via $\varGamma_i=\sum_{j\neq i}^{N}\lVert\triangle w_t^{i''}-\triangle w_t^{j''}\rVert$
		\ENDFOR
		\STATE Let $\varGamma=[\varGamma_1,...,\varGamma_N]$
		\STATE Let $\varGamma^{'}=\varGamma/{\rm median}(\varGamma)=[\varGamma_1^{'},...,\varGamma_N^{'}]$
		\FOR {client $i=1,...,N$}
		\IF {$\varGamma_i^{'}>\varepsilon_1$}
		\STATE Exclude gradients of client $i$ based on threshold $\varepsilon_1$  \textcolor{blue}{//Only valid in this round}
		\ENDIF
		\ENDFOR
		\STATE After excluding satisfied gradients of clients, client set $[N]$ becomes $[\tilde{N}]$
		\STATE Cluster $\{\triangle w_t^{i''}:i \in [\tilde{N}]\}$ using Kmeans algorithm, and get $C$ clusters
		\FOR {Cluster $c=1,...,C$ in parallel}
		\FOR {client $j$ in cluster $c$}
		\STATE Let $s_{cj}$ be the similarity set between client $j$ and the other clients in cluster $c$   \textcolor{blue}{//Use the flattened gradients $\triangle w_t^{j'}$ to compute cosine similarity}
		\STATE Let $v_{cj}={\rm average}(s_{cj})$
		\ENDFOR
		\STATE Let $v_c$ be the set of $v_{cj}$ in cluster $c$
		\STATE Let $v_{cmed}={\rm median}(v_c)$
		\ENDFOR
		\STATE Let $v_{med}$ be the set of $v_{cmed}$
		\STATE The final selected cluster is $c={\rm argmin}(v_{med})$
		\STATE Exclude the unsatisfied gradients of clients in selected cluster $c$ like line $15-24$ with threshold $\varepsilon_2$
		\STATE The final selected clients set denotes as $[\tilde{N}^{'}]$
		\STATE Update global model $w_t=w_{t-1}+\sum_{i\in [\tilde{N}^{'}]}p_i\triangle w_t^i$
		\ENDFOR
		\STATE Return final global model $w_T$
	\end{algorithmic}
\end{algorithm}

\section{Methodology}
In this section, we introduce the proposed robust aggregation oracle \solution \ in Algorithm \ref{alg:\solution}. In contrast to the existing majority-based algorithms, our algorithm \solution \ is intended for a FL setting where the central server can only access to the gradients of clients under IID and Non-IID scenarios, and does not assume that the backdoored clients is less than the benign clients. In addition to this, \solution \ require no auxiliary information outside of the learning process.

Our algorithm \solution \ has two key insights: one is the gradients pushed by backdoored clients and benign clients can be distinguished using PCA and partitioned into different clusters using Kmeans clustering algorithm, and another is the gradients of backdoored clients appear more similar to each other than benign clients. \solution \ uses the two insights to effectively select the benign gradients to aggregate, so as to fully avoid the participation of the backdoor clients.

\textbf{\solution \ design.} On the client side, during the training process, at round $t=1,...,T$, the clients download the latest model and update their models with $E$ iterations and datasets $D^{'}$, and then send the updated gradients to the central server. On the central server side, and the central server collects the gradients pushed by clients in synchronous rounds, and performs aggregation to generate the global model for next round training using \solution.

Figure \ref{flow} shows the process of \solution \ on the central server side, where the gray dotted line is the main work of \solution. The general process of \solution \ is: first, reduce the dimension of gradients using PCA; second, exclude the outliers which are harmful to the next process; third, cluster the gradients set using Kmeans clustering method; fourth, select the optimal gradients based on cosine similarity; fifth, exclude the outliers again; finally, aggregate the selected gradients to the global model for next round training. To clearly clarify how \solution \ works and how to solve above mentioned challenges on the central server side at round $t$, we will describe \solution \ in detail in Algorithm \ref{alg:\solution}.

\begin{figure*}[!h]
	\centering
	\includegraphics[width=6in]{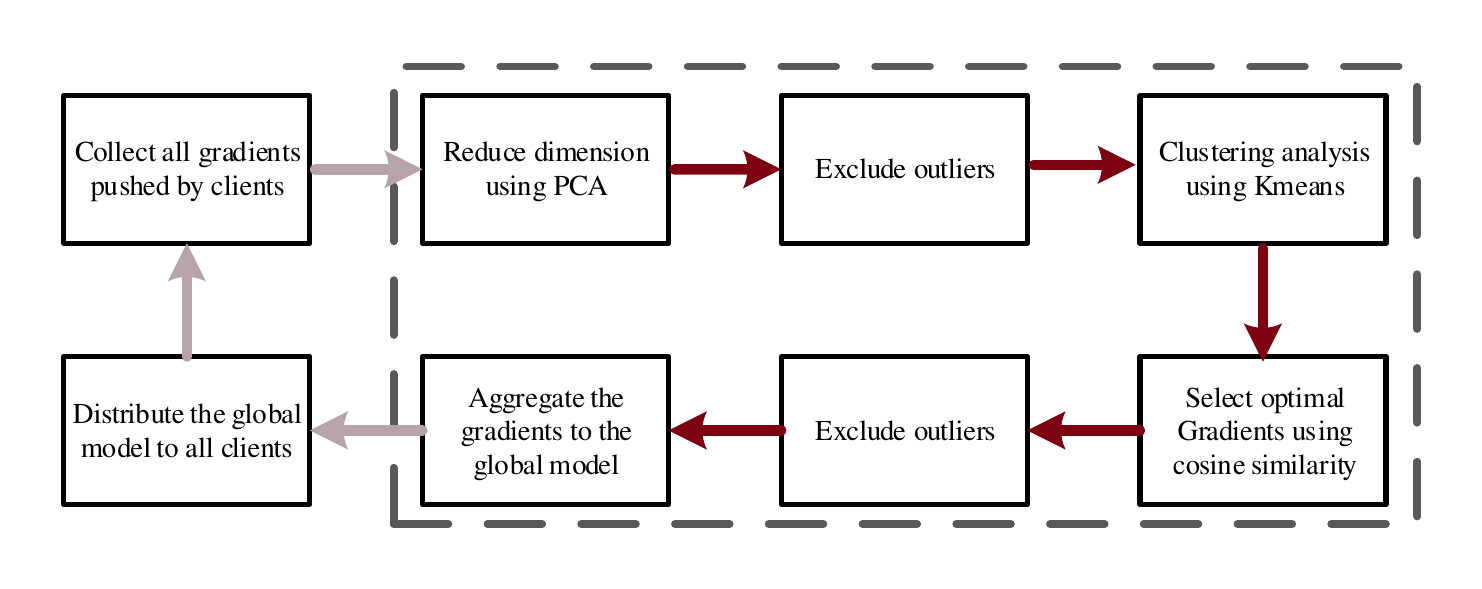}
	\caption{The flow of our algorithm \solution \ on the central server side.}
	\label{flow}
\end{figure*}


\textbf{Step 1.} The central server collect all gradients $\{\triangle w_t^i:i \in [N]\}$ pushed by backdoored clients and benign clients (line 12).

\textbf{Step 2.} Flatten the gradient of every client and get a gradient set $\{\triangle w_t^{i'}:i \in [N]\}$ with $N \times m$ dimensions, where $N$ and $m$ represent the number of clients and model parameters respectively (line 13). Then, reduce the dimension of $\{\triangle w_t^{i'}:i \in [N]\}$ using PCA, and get a new gradient set $\{\triangle w_t^{i''}:i \in [N]\}$ with $N\times h$ dimensions. Here, we set $h$ to be 2 (line 14).

\textbf{Step 3.} Calculate the sum of Euclidean distance between each client $i$ and the other clients like $\varGamma_i=\sum_{j\neq i}^{N}\lVert\triangle w_t^{i''}-\triangle w_t^{j''}\rVert$ (line 16). Then, the sum of distance per client is dimensionless based on the median value (line 19). Exclude the outliers lager than threshold $\varepsilon_1$ (line 22). Here, we set $\varepsilon_1$ to be 10. Note that we use the gradients after dimensionality-reduction to compute Euclidean distance. This step addresses \textbf{challenge 2}.

\textbf{Step 4.} Use Kmeans clustering algorithm to cluster $\{\triangle w_t^{i''}:i \in [\tilde{N}]\}$, and get $C$ clusters, where $[\tilde{N}]$ represents the client set after removing unsatisfied clients from original client set $[N]$ in \textbf{Step 3} (line 26).

\textbf{Step 5.} Select the optimal cluster for aggregation based on similarity. In \solution \,  we use cosine similarity to measure the angular distance between client $i$ and client $j$ such that $s_{ij}=\emph{cosine\_similarity}(\triangle w^i,\triangle w^j)$. Specially, in each cluster $c$, computer the cosine similarity between each client $j$ and the others, and $s_{cj}$ denotes the similarity set of client $j$ in cluster $c$ (line 29). Then, take an average for each client's similarity set $s_{cj}$, denote as $v_{cj}$ (line 30), and each client's $v_{cj}$ constitutes set $v_c$ (line 32). Choose the median of $v_c$ as the overall similarity of cluster $c$ (line 33), denote as $v_{cmed}$. Each cluster's similarity $v_{cmed}$ constitutes set $v_{med}$ (line 35), finally select cluster $c$ with the minimum value in $v_{med}$ (line 36). Note that we utilize the original gradients to compute the similarity, not the gradients after dimensionality-reduction. This step addresses \textbf{Challenge 3}.

\textbf{Step 6.} Though we have selected an effective cluster for aggregation, a few backdoored clients may be still present in the selected cluster $c$. Therefore, similar to \textbf{Step 3}, remove satisfied clients based on $\varepsilon_2$ in the selected cluster $c$ (line 37). Here, to completely exclude backdoored clients that may be in the selected cluster $c$, we set $\varepsilon_2$ to be 4 smaller than $\varepsilon_2$.

\textbf{Step 7.} Use FedAvg algorithm to aggregate the selected gradients such that $w_t=w_{t-1}+\sum_{i\in [\tilde{N}^{'}]}p_i\triangle w_t^i$, where $[\tilde{N}^{'}]$ represents the selected client set (line 39).

\textbf{Step 8.} The central server sends updated global model to all clients for next round training (line 11).

\textbf{Step 9.} After multiple rounds training, the central server generate the final global model $w_T$ (line 41).

\textbf{Convergence analysis.} Note that our algorithm \solution \ actually selects partial clients from the client set to aggregate the global model based on FedAvg. Li \cite{fedavg} proves that FedAvg algorithm is convergent using partial clients under Non-IID scenario. \solution \ thus shares the same convergence property as the FedAvg.

\section{Experiments}
In this section, we evaluate the performance of \solution \ on image classification and sentiment analysis tasks with three common machine learning models over three well-known public datasets under various scenarios. We test our approach \solution's effectiveness by comparing to four baseline aggregation algorithms. Our experiments are implemented using PyTorch framework with about 3000 lines code.

\subsection{Datasets and Models}
We demonstrate three public datasets and three well-known machine learning models. For image classification task, we use MNIST\cite{2016Communication} and FEMNIST \cite{caldas2018leaf} datasets. For the sentiment analysis task, we use Amazon Fine Food Reviews from kaggle \cite{Amazon}. The details of all three datasets and machine learning models are as follows, and Table \ref{tab:1} describes the experimental datasets and machine learning models.

\textbf{MNIST.} MNIST dataset \cite{2016Communication} contains 60000 training images and 10000 testing images with 10 labels, and the size of each image is 28$\times$28. In IID scenario, all training samples are uniformly divided into 100 parties, each client holds a party which consists of 600 training samples with 10 labels. To implement Non-IID scenario, a Dirichlet distribution is used to divide training images into 100 parties, each party holding samples which are different from other parties in both class and quantity.

\textbf{FEMNIST.} FEMNIST dataset \cite{caldas2018leaf} includes 801074 samples with 62 object classes distributed among 3500 writers, and the size of each image is also $28\times28$. Each writer represents a client, resulting in a heterogeneous federated scenario.

\textbf{Amazon.} We download Amazon Fine Food Reviews from kaggle \cite{Amazon}, and it contains 568454 food reviews with 1-5 score as 5 object classes. We only sample 20000 reviews of every class to train a model, total 100000 data samples. Similar to MNIST, we split the training data for 100 parties in IID and Non-IID manners. 80\% of data samples are used for training and the rest for testing.

\textbf{Models.} For MNIST dataset, we train a multi-class logistic regression (LR) model which contains one softmax layer with 784 units, total 7850 parameters. For FEMNIST dataset, we train a convolutional neural network (CNN) model with 2 CNN layers (7$\times$7$\times$32 and 3$\times$3$\times$64), followed by a fully connected (FC) layer with 3136 units. There are total 0.2M parameters in CNN model. For Amazon dataset, we train a one layer directional long short-term memory (LSTM) model with 100 hidden units, followed by 1 FC layers with 64 units. We first embed 56785 words into 100 dimensions, and each sample has a maximum of 100 words, and less than 100 words are padded with 0. There are total 6M parameters in LSTM model. The datasets and models are summarized as Table \ref{tab:1}.

\begin{table}[!h]
	\caption{The description of datasets and models}
	\label{tab:1}       
	\begin{center}
		\begin{tabular}{lcccc}
			\toprule
			
			
			Dataset & Classes & Data size  & Features & Model  \\
			
			\noalign{\smallskip}\hline\noalign{\smallskip}
			
			MNIST & 10 & 60000 & 784 & LR \\
			
			FEMNIST & 62 & 801074 & 784 & CNN  \\
			
			Amazon & 5 & 80000 & 100 & LSTM \\
			
			\bottomrule
		\end{tabular}
	\end{center}  	
\end{table}

\subsection{Experiment Setup}
We use PS structure which consists a central server and multiple clients to conduct our experiments until convergence. The training process has been illustrated in Section \ref{sec:Introduction}. Next, we will expound backdoor attack patterns for two tasks and various experiment scenarios.

\textbf{Backdoor attack.} The goal of backdoor attack is to change the global model's behavior on some data samples with certain backdoor triggers, while maintain high performance on normal data samples. For image classification task,  we consider some certain pixels as backdoor triggers. Specially, the backdoored client poisons its local training data samples using backdoor patterns in lower left corner of Figure \ref{fig:image} and swaps the original label into label ``0''. For sentiment analysis task, the backdoored client inserts backdoor \textit{``The weather is so good, I want to eat noodles.''} to its local training data, and swaps original label into label ``0'', as illustrated in Figure \ref{fig:sentiment}. The backdoored attackers enforce the model to classify the test data with certain backdoor as a certain label they pre-config.

\begin{figure}[tbp]
	\centering
	\subfigure[Image's backoor]{
		\includegraphics[width=1.5in]{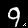}
		\label{fig:image}}
	\subfigure[sentiment's backoor]{
		\raisebox{1.5\height}{\includegraphics[width=1.5in]{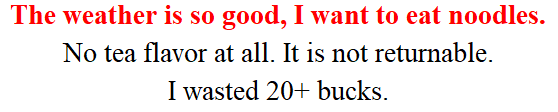}}
		\label{fig:sentiment}}
	\caption{The backdoor triggers in image and sentiment.}
	\label{fig:backdoor}
\end{figure}

\textbf{Different number of attackers (DNA).}  In each round, we randomly select 100 clients to train a model, where a certain number of clients are backdoored attackers. We evaluate our algorithm \solution \ under both IID and Non-IID scenarios with 10\%, 50\% and 90\% backdoored clients, respectively. We verify that the superiority of \solution \ compared to four state-of-the-art robust algorithms under different number of attackers.

\textbf{Different Non-IID scenarios (DNS).} Due to the differences in class and data size, there are various Non-IID scenarios. To evaluate the effectiveness of \solution \ under different Non-IID scenarios, we use different datasets to realize different Non-IID scenarios. In addition to this, we also simulate different Non-IID scenarios through changing the parameter $\alpha$ of Dirichlet distribution on MNIST. Parameter $\alpha$ increases from 0.1 to 2 represent a variety of Non-IID scenarios.

\textbf{Different number of clients (DNC).} We also evaluate the effectiveness of \solution \ in the face of different number of clients. In FEMNIST dataset, we respectively select 50, 100, 200, 400, 600, 800, 1600 clients among 3500 clients to conduct a series of experiments.

\textbf{Distributed backdoor attack (DBA).} Xie \cite{xie2019dba} proves that the distributed backdoor attack (DBA) is much stronger than centralized backdoor attack, and can beat most of state-of-the-art robust algorithms. Similar to \cite{xie2019dba}, the centralized backdoor patterns are divided into four distributed backdoor pattern. In our experiment, we poison 40 clients among 100 total clients, and every 10 clients as a group performs each local backdoor attack.

To be fair, the only difference between benign clients and backdoored clients is the training data. Table \ref{tab:2} summarizes the total experiment scenarios.

\begin{table*}[]
	\caption{The description of experiment scenarios}
	\label{tab:2}       
	\begin{center}
		\begin{tabular}{ccc}
			\toprule
			
			
			Scenario & Description & Dataset \\
			
			\noalign{\smallskip}\hline\noalign{\smallskip}
			
			DNA & Poisoning 10\%, 50\%, 90\% clients & All  \\
			
			DNS & Changing Dirichlet distribution $\alpha$ from 0.1 to 2 & MNIST   \\
			
			DNC & 50, 100, 200, 400, 600, 800, 1600 clients & FEMNIST  \\
			
			DBA & 40 poisoned clients, every 10 clients performing local backdoor attack & All \\
			
			\bottomrule
		\end{tabular}
	\end{center}  	
\end{table*}

\subsection{Comparison Algorithms}
We compare our algorithm \solution \ with four typical robust aggregation algorithms: FoolsGold Multi-Krum, GeoMed and RFA.
\textbf{FoolsGold.} FoolsGold \cite{2018Mitigating} reduces aggregation weights of participating parties that repeatedly contribute similar gradient updates while retaining the weights of parities that provide different gradient updates.

\textbf{Multi-Krum.} Multi-Krum \cite{blanchard2017machine} calculates the total Euclidean distance from the $n-f-2$ nearest neighbors for each local update. The $f$ local updates with the highest distances are excluded and average the rest of local updates. Nevertheless, Multi-Krum relies on $f$ parameter, and prior knowledge of $f$ is an unrealistic assumption when defending against backdoor attack.

\textbf{GeoMed.} GeoMed \cite{chen2018distributed} generates a global model update using the geometric median of the local model updates, including the local model updates pushed by backdoored clients.

\textbf{RFA.} Similar to GeoMed, RFA \cite{pillutla2019robust} aggregates a global model update and appear robust to outliers by replacing the weighted arithmetic mean with an approximate geometric median, so as to reduce the impact of the contaminated updates.

\subsection{Experiment Results}
\subsubsection{The Experiment Results of DNA}
For all datasets,  we run four existing typical robust algorithms FoolGold, Multi-Krum, GeoMed, RFA and our proposed \solution \, and train three above machine learning models until convergence. In MNIST and Amazon datasets, we consider both IID and Non-IID scenarios, while only consider Non-IID scenario for FEMNIST due to its inherently heterogeneous.

\textbf{MNIST under IID scenario.} For MNIST under IID scenario, Figure \ref{fig:iidmnist} shows the performance of the four typical robust aggregation algorithms and \solution \ with an increasing number of backdoored clients: \solution \ outperforms the other four existing algorithms. The four existing algorithms and \solution \ can reach a fine training accuracy on normal testing samples, while the performance of the five algorithms facing backdoor attack is quite different.

\begin{figure*}[]
	\centering
	\subfigure[10\% backdoored clients]{
		\includegraphics[width=2.22in]{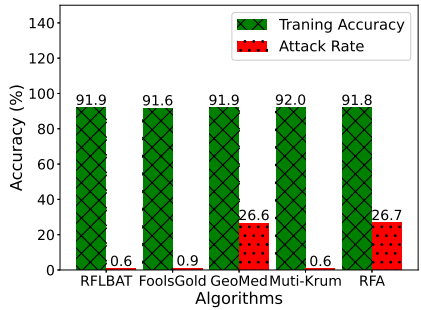}
		\label{iidmnist:10}}
	\subfigure[50\% backdoored clients]{
		\includegraphics[width=2.22in]{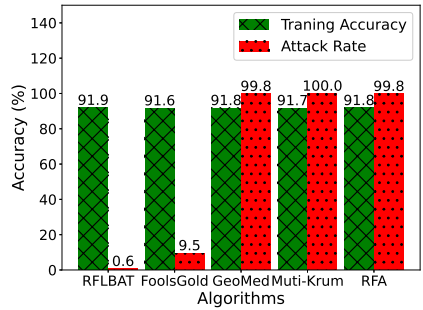}
		\label{iidmnist:50}}
	\subfigure[90\% backdoored clients]{
		\includegraphics[width=2.22in]{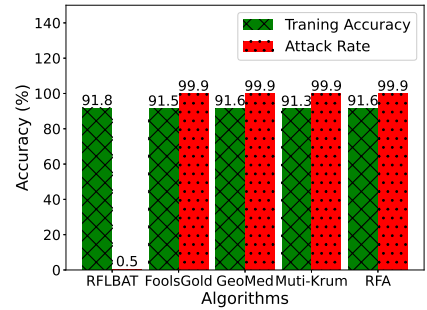}
		\label{iidmnist:90}}
	\caption{The performance of Multi-Krum, GeoMed, RFA algorithms and \solution \ on MNIST under IID scenario with different number of backdoored clients.}
	\label{fig:iidmnist}
\end{figure*}
\begin{figure*}[]
	\centering
	\subfigure[10\% backdoored clients]{
		\includegraphics[width=2.22in]{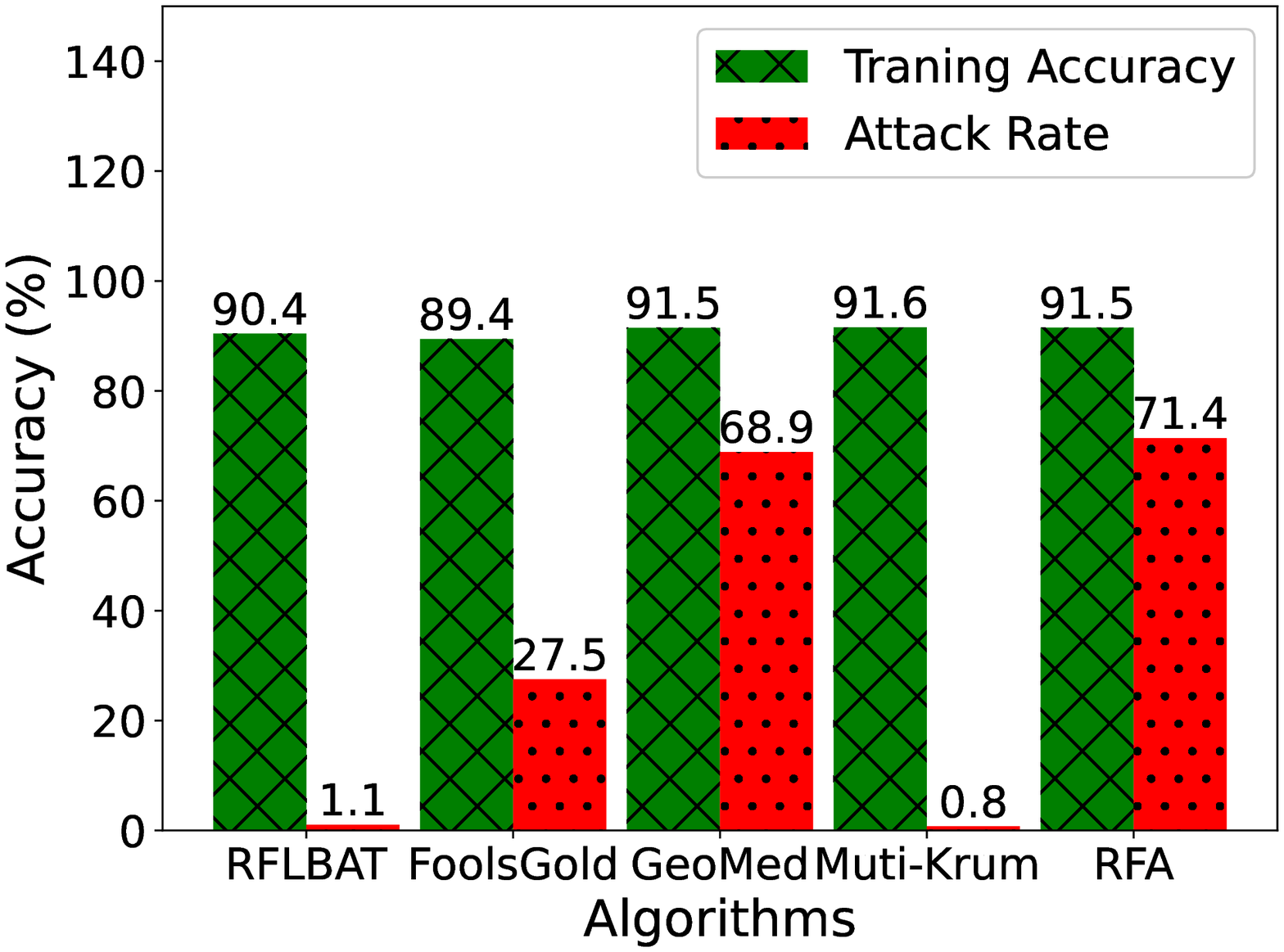}
		\label{noniidmnist:10}}
	\subfigure[50\% backdoored clients]{
		\includegraphics[width=2.22in]{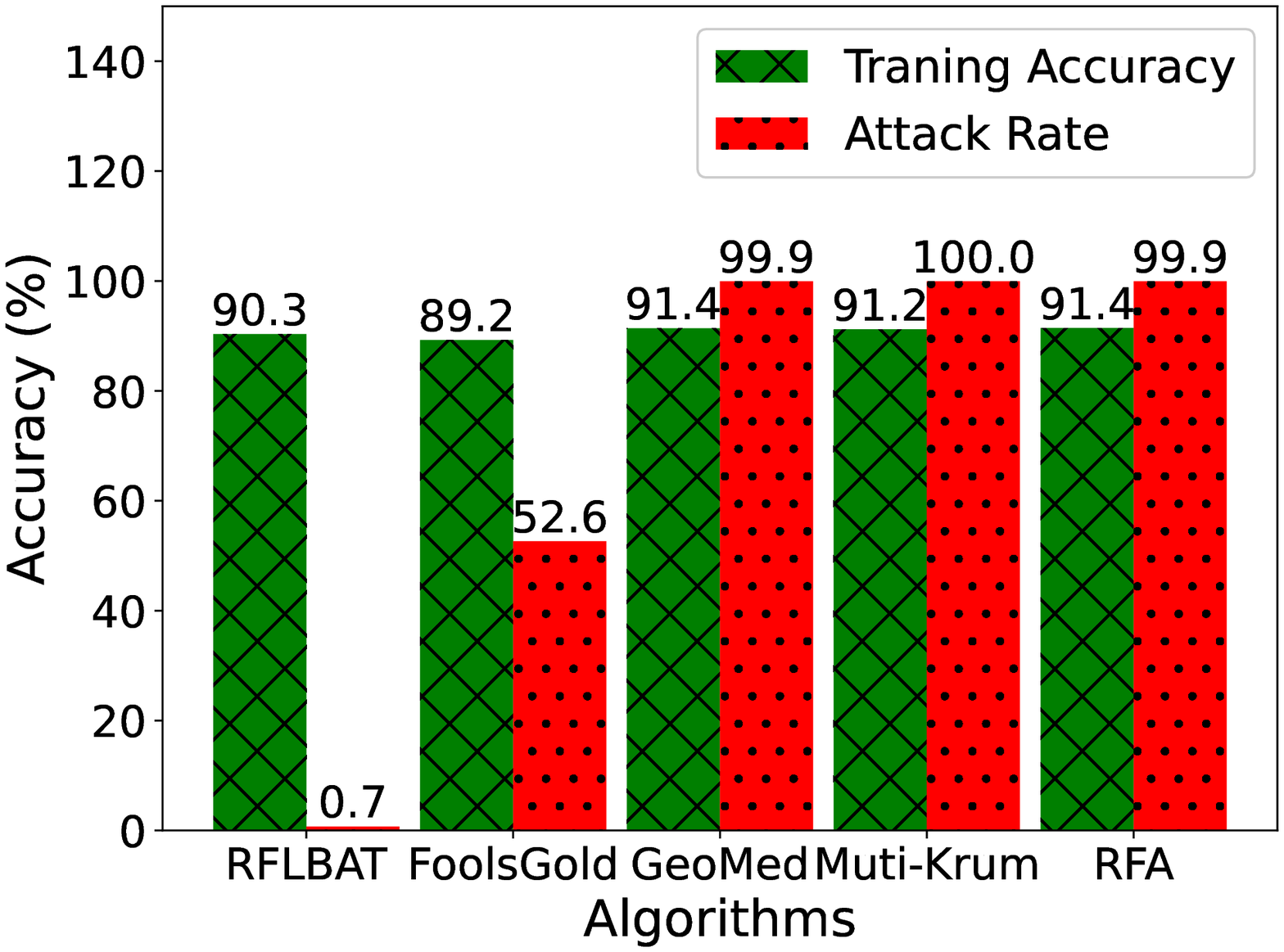}
		\label{noniidmnist:50}}
	\subfigure[90\% backdoored clients]{
		\includegraphics[width=2.22in]{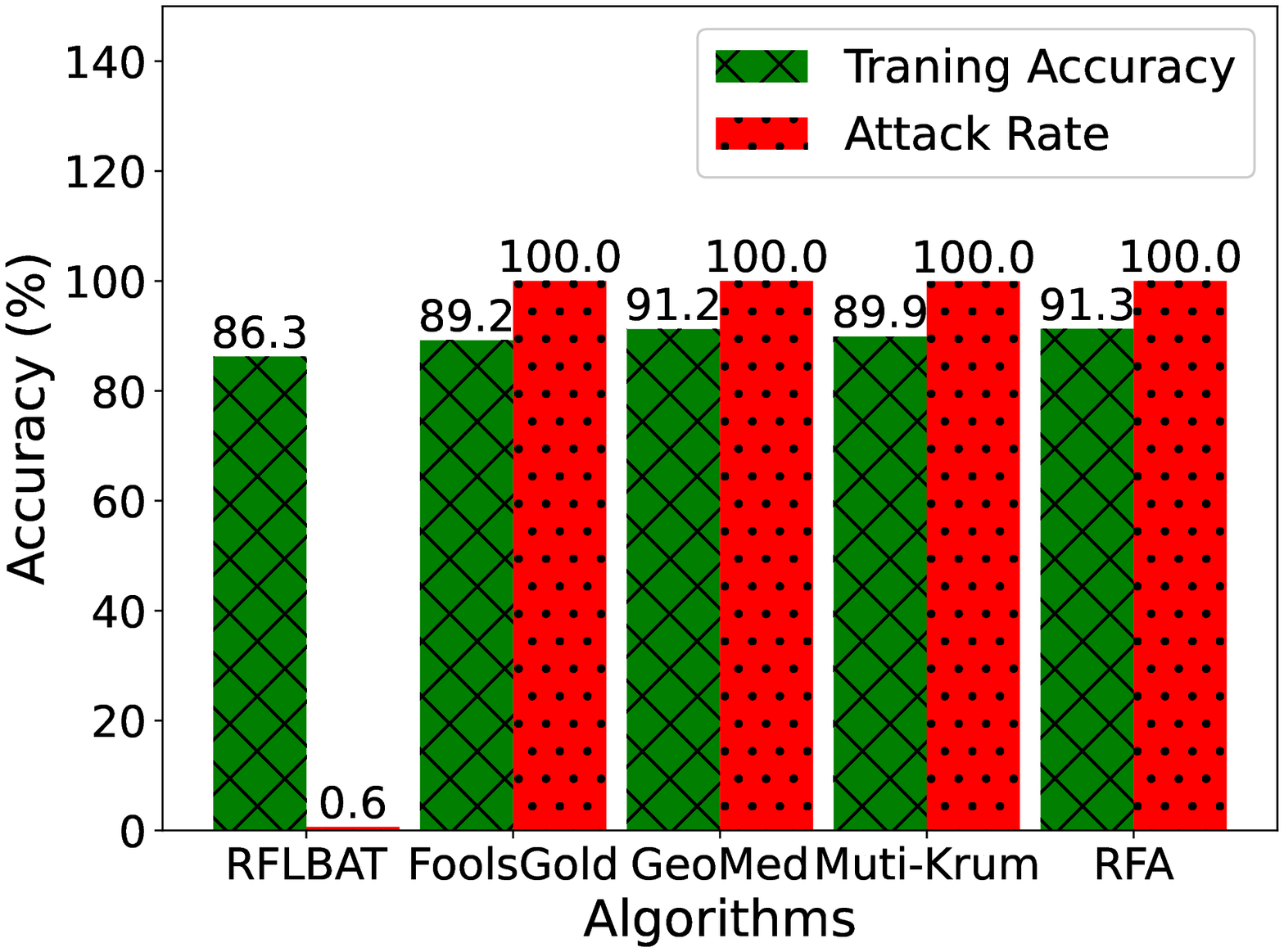}
		\label{noniidmnist:90}}
	\caption{The performance of Multi-Krum, GeoMed, RFA algorithms and \solution \ on MNIST under Non-IID scenario with different number of backdoored clients.}
	\label{fig:noniidmnist}
\end{figure*}
\begin{figure*}[]
	\centering
	\subfigure[10\% backdoored clients]{
		\includegraphics[width=2.22in]{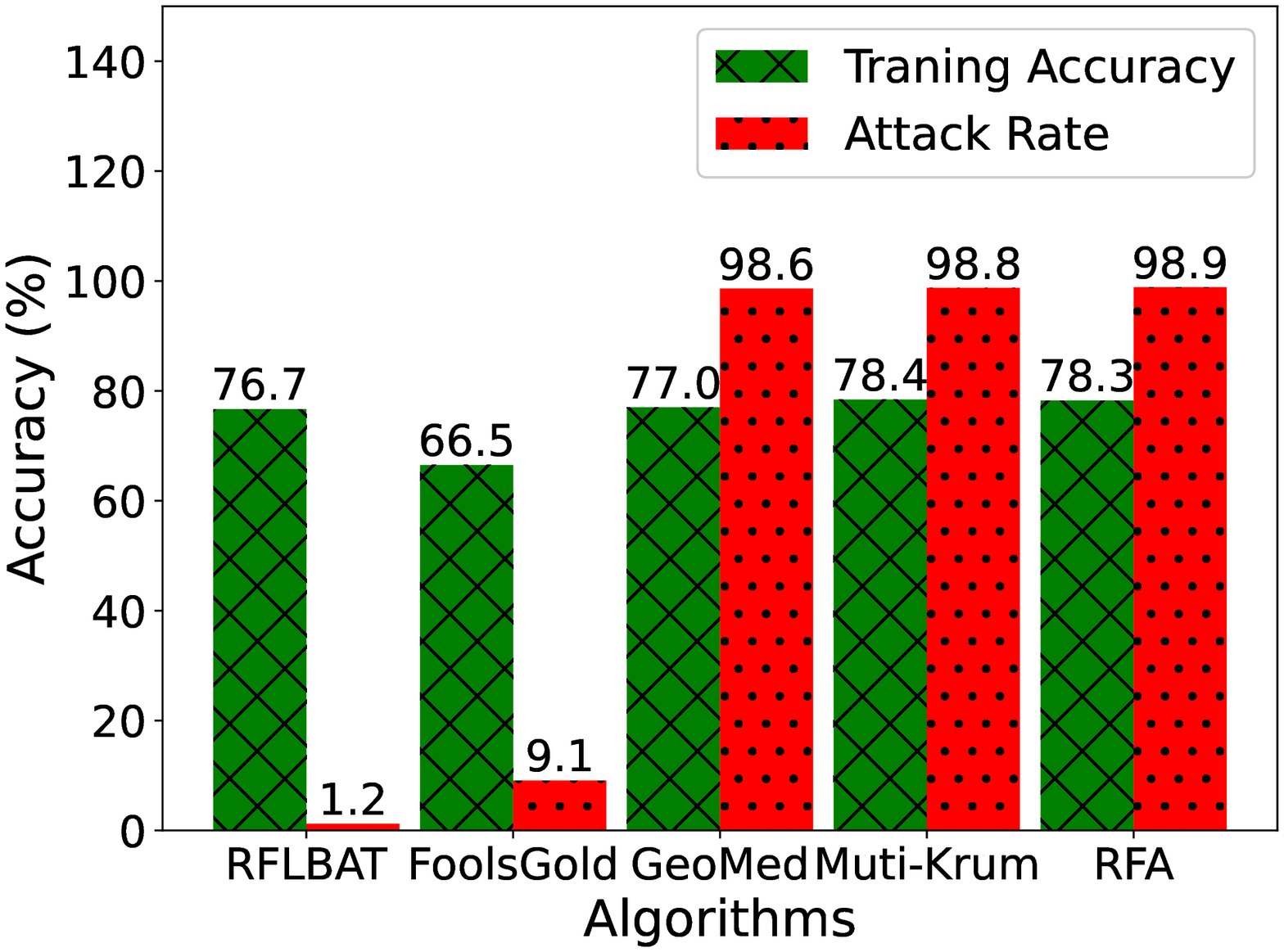}
		\label{femnist:10}}
	\subfigure[50\% backdoored clients]{
		\includegraphics[width=2.22in]{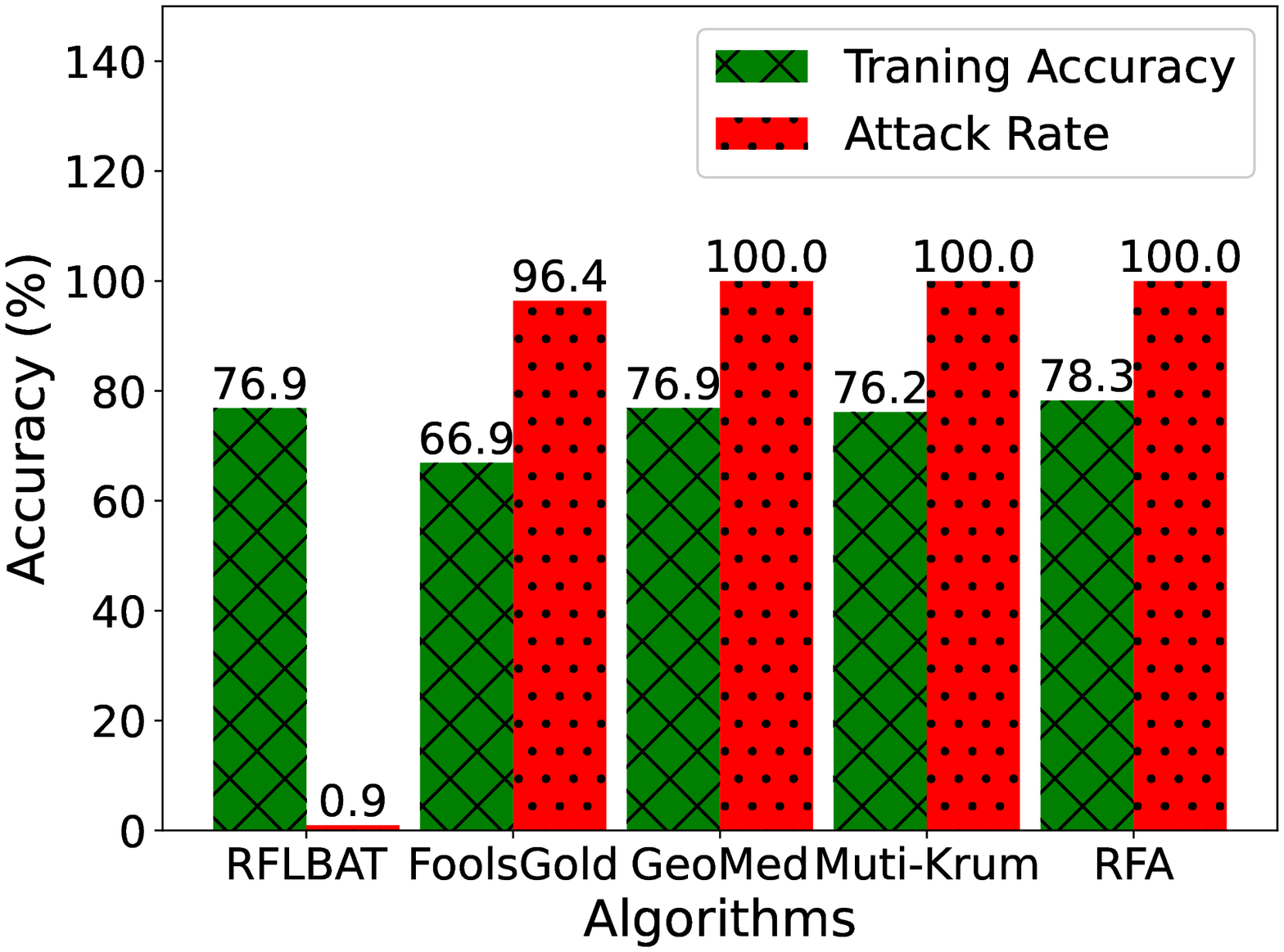}
		\label{femnist:50}}
	\subfigure[90\% backdoored clients]{
		\includegraphics[width=2.22in]{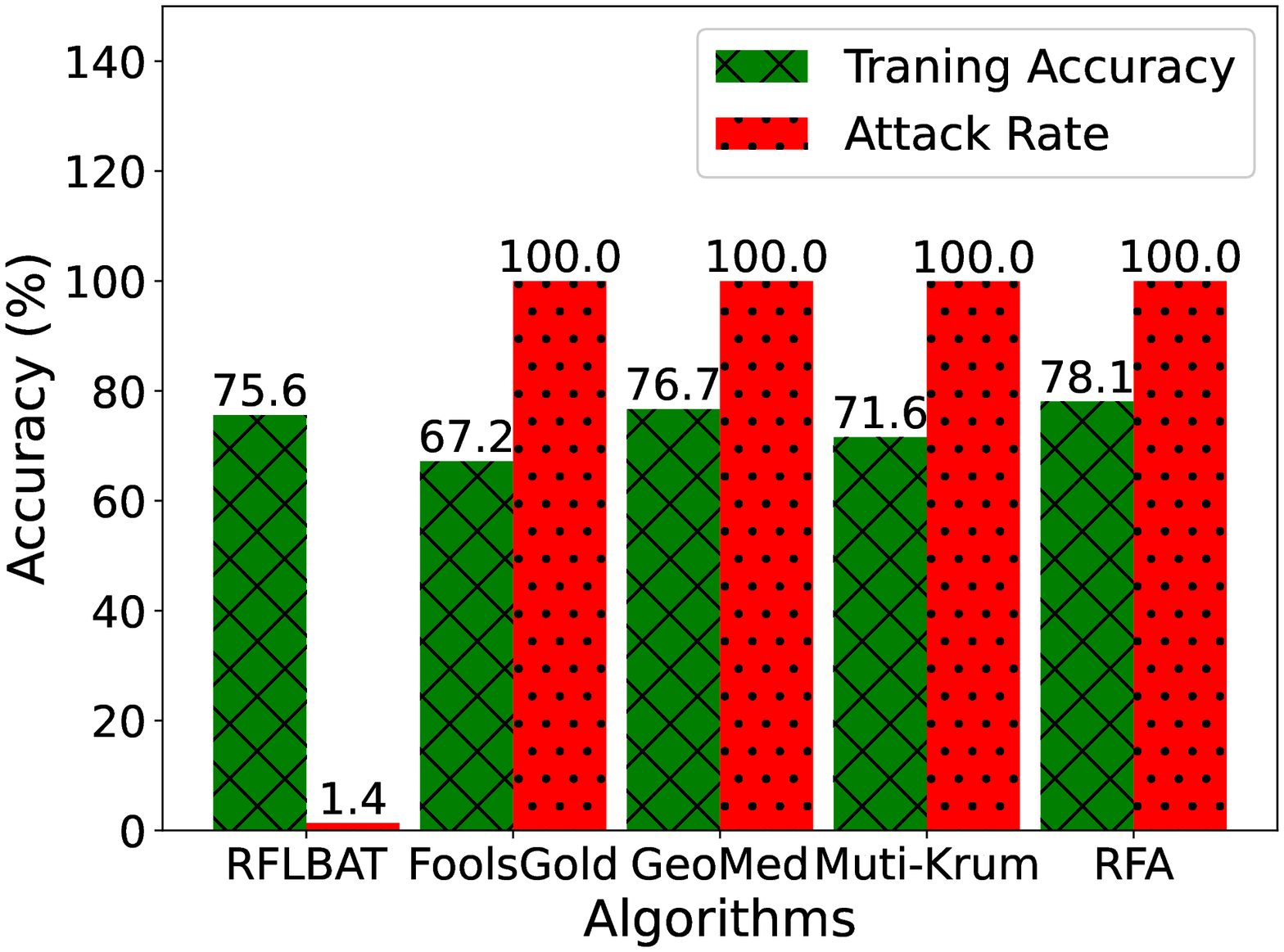}
		\label{femnist:90}}
	\caption{{\color{blue}The performance of Multi-Krum, GeoMed, RFA algorithms and \solution \ on FEMNIST under Non-IID scenario with different number of backdoored clients.}}
	\label{fig:femnist}
\end{figure*}
\begin{figure*}[]
	\centering
	\subfigure[10\% backdoored clients]{
		\includegraphics[width=2in]{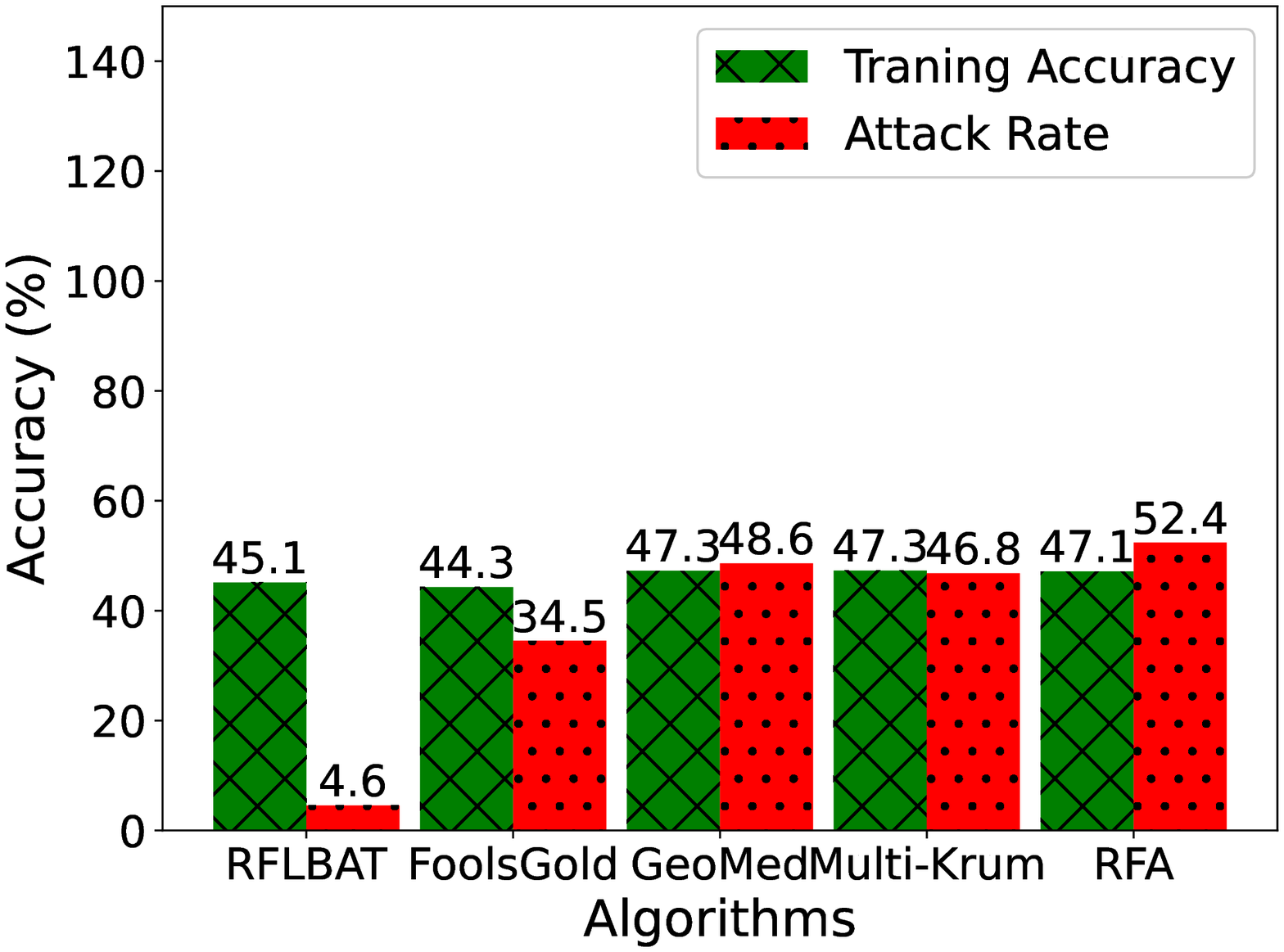}
		\label{sentiment:10}}
	\subfigure[50\% backdoored clients]{
		\includegraphics[width=2in]{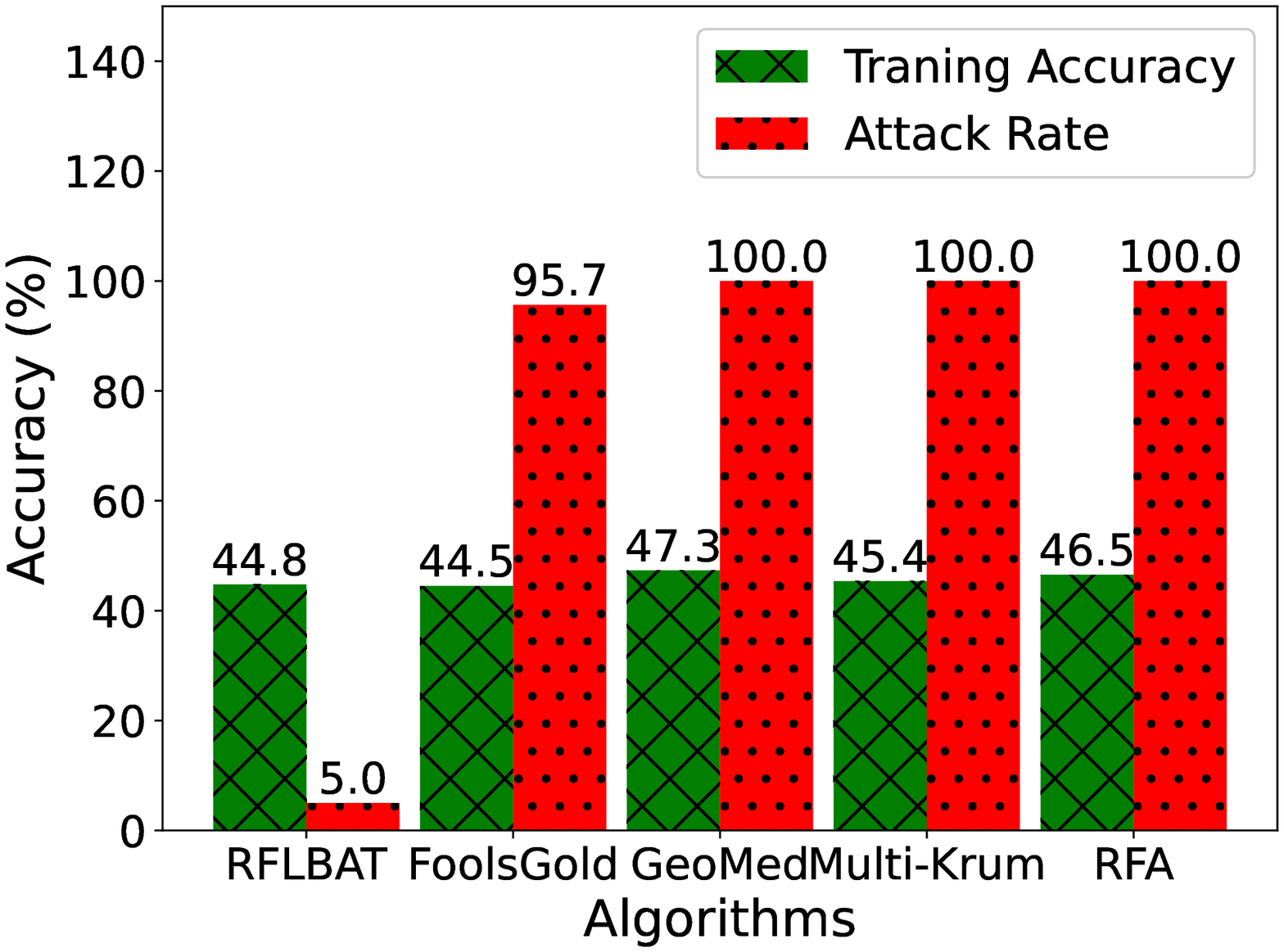}
		\label{sentiment:50}}
	\subfigure[90\% backdoored clients]{
		\includegraphics[width=2in]{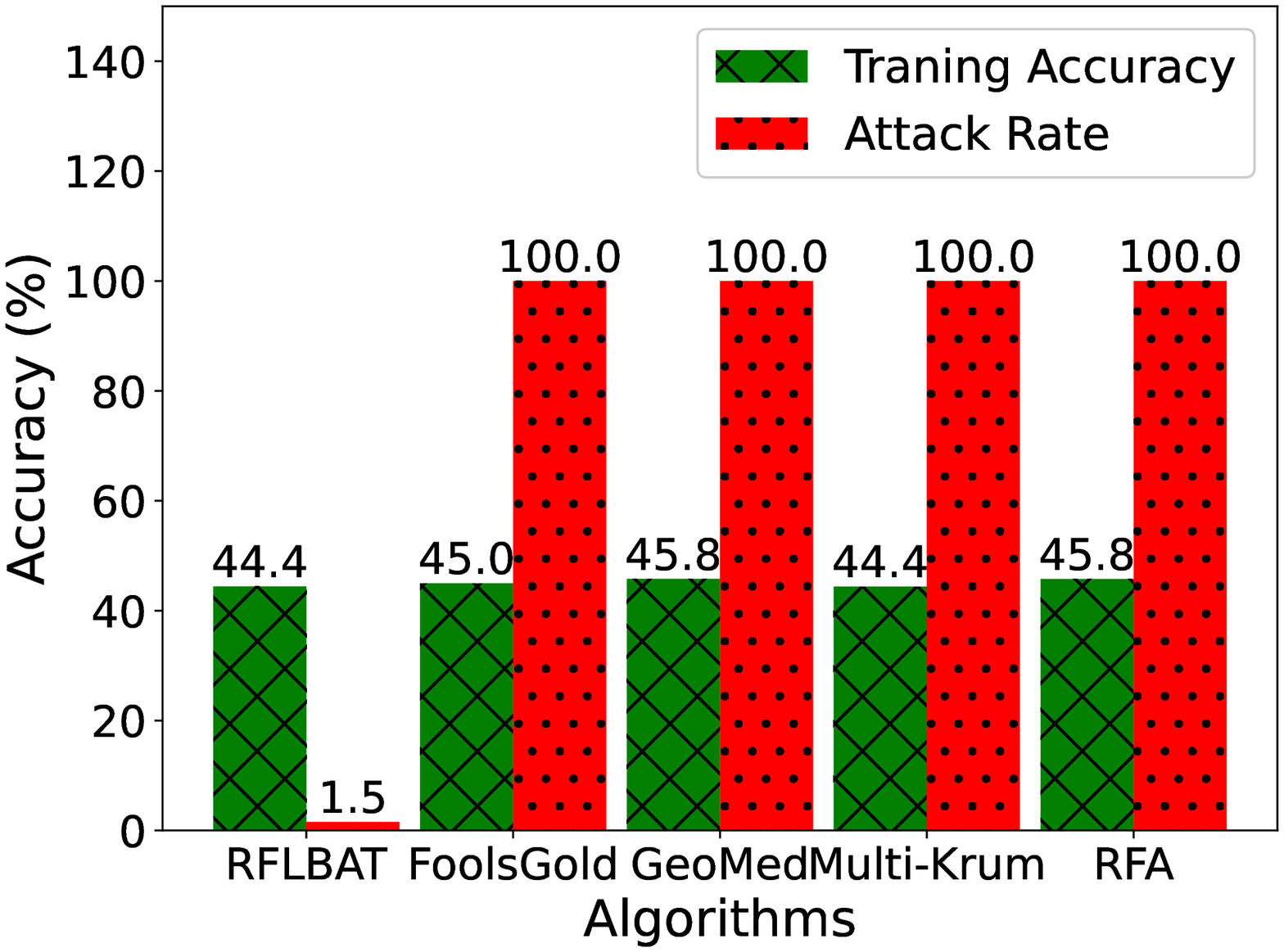}
		\label{sentiment:90}}
	\caption{The performance of FoolsGold, Multi-Krum, GeoMed, RFA algorithms and \solution \ on Amazon under IID scenario with different number of backdoored clients.}
	\label{fig:sentiment_iid}
\end{figure*}
Specially, with the increasing number of backdoored clients, the four existing typical algorithms are gradually failed against backdoor attack, but \solution \ always works. When the number of backdoored clients is 10\%, as shown in Figure \ref{iidmnist:10}, FoolsGold, Multi-Krum and \solution \ can effectively prevent this attack while maintaining high training accuracy. It indicates that the three algorithms can completely nullify all backdoored gradients during aggregating the global model. Compared to FoolsGold, Multi-Krum and \solution \, GeoMed and RFA perform worst, and the attack rate can reach about 26\%, which means the backdoor triggers have been moderately inserted into the global model. Because both GeoMed and RFA are based on geometric median, the attack rate is similar.

As the number of backdoored clients increases to 50\%, Figure \ref{iidmnist:50} shows the results of five algorithms against this attack. We can see that \solution \ can also resist this attack and meanwhile maintain high training accuracy. Although FoolsGold can defend against this attack, the attack rate of using FoolsGold reaches 9.5\% which is higher than \solution \. This demonstrates that the performance of FoolsGold is worse than \solution. By contrast, Multi-Krum, GeoMed and RFA can hardly defend against this attack, resulting in reaching nearly 100\% attack rate. It indicates that the backdoor triggers have fully been inserted into the global model using the three algorithms.

As the four existing typical algorithms and \solution \ face lager groups of backdoored clients, for example 90\% backdoored clients scenario, the results are shown in Figure \ref{iidmnist:90}. \solution \ still exhibits robust to this attack, and the attack rate is nearly 0.5\% which may be caused by the model error. In contrast, FoolsGold, Multi-Krum, GeoMed and RFA perform worst against this attack, and the attack rate reaches nearly 100\%. It means that the three algorithms are completely ineffective against this attack.

\textbf{MNIST under Non-IID scenario.} Figure \ref{fig:noniidmnist} shows the performance of FoolsGold, Multi-Krum, GeoMed, RFA and \solution \ on MNIST under Non-IID setting with an increasing number of backdoored clients: \solution \ outperforms the other four existing typical algorithms against these backdoor attacks. Compared with Figure \ref{fig:iidmnist},  we can find the similar conclusions in Figure \ref{fig:noniidmnist}. Nevertheless, there are some differences between IID and Non-IID scenarios.

To be specific, as the number of backdoored clients is 10\%, \solution \ and Multi-Krum can also defend against this attack, but the attack rate reaches to nearly 27.5\%,68.9\% and 71.4\% using FoolsGold, GeoMed and RFA respectively, which is different from Figure \ref{iidmnist:10}. Obviously, it is relevant to the data distribution. In IID scenario, the training data of each client is uniformly sampled from the whole data set and every client shares the common data size and data classes while every client shares the little or no data in Non-IID scenario. The model update are not only affected by data features, but also by the data size and data classes, which will seriously degrade the performance of FoolsGold, GeoMed and RFA.

As the number of backdoored clients increases to 50\% and 90\%, similar to IID scenario, \solution \ are still robust with these attacks, while FoolsGold, Multi-Krum, GeoMed and RFA algorithms are disabled. This demonstrates that \solution \ can remove all malicious gradients even in the Non-IID scenario. It is worth noting that although \solution \ get high training accuracy, it is slightly lower than the other four algorithms in 90\% backdoored clients scenario. The reason is simple: \solution \ tries to select the best cluster containing the most benign gradients to aggregate, whereas the best cluster may not include all benign gradients in Non-IID scenario. Therefore, \solution \ loses some normal gradients from unselected clients, consequently resulting in slightly lower training accuracy than the four contrastive algorithms.


\textbf{FEMNIST under Non-IID scenario.}Figure \ref{fig:femnist} shows the performance of FoolsGold, Multi-Krum, GeoMed, RFA and \solution \ on FEMNIST under Non-IID scenario with an increasing number of backdoored clients. Similar to Figure \ref{fig:iidmnist} and Figure \ref{fig:noniidmnist}, as the number of backdoored clients grows, \solution \ can still effectively defend against backdoor attacks at the cost of losing less than 3\% of the training accuracy, while Multi-Krum, GeoMed and RFA are completely ineffective even if the number of backdoored clients is 10\%, and FoolsGold gets 9.1\% attack rate which is worse than that of \solution. Different from MNIST under 10\% backdoored clients scenario, the attack rate of using Multi-Krum, GeoMed and RFA can reach nearly 99\%. This indicates that FEMNIST is more heterogeneous than MNIST in our experiment settings. When the number of backdoored clients increases to 50\% and 90\%, the four existing algorithms are completely disabled.  It should be noted that the training accuracy of FoolsGold is the worst among the five algorithms, which is different from Figure \ref{fig:noniidmnist}. It may be relevant to the peculiarity of FoolsGold which  reduces aggregation weights of participating parties based on similarity. During aggregation, FoolsGold may severely reduce the weights of benign clients, resulting in degrading the contribution of benign clients.

\textbf{Amazon under IID scenario.} Figure \ref{fig:sentiment_iid} shows the performance of FoolsGold, Multi-Krum, GeoMed, RFA and \solution \ on Amazon dataset. For sentiment classification, there are no big differences compared to image classification.  We can also draw a conclusion that \solution \ is effectively defend against backdoor attack and outperforms the four existing typical algorithms. We should pay attention to the attack rate using \solution: the attack rate of 90\% backdoored clients is higher than that of 10\% backdoored clients and 50\% backdoored clients, which goes against our intuitive assumptions. The reason for this result may be that the number of backdoored clients participating in the aggregation under 10\% backdoored clients and 50\% bacdoored clients is more than that under 90\% backdoored clients. This also suggests that the intensity of backdoor attack is not necessarily related to the number of backdoored attackers using \solution.

\textbf{Amazon under Non-IID scenario.} In Amazon under IID scenario, all robust aggregation algorithms are failed except \solution. For Amazon dataset under Non-IID, we only evaluate the performance of the five algorithms against 90\% backdoored clients, and the result is shown in Figure \ref{sentimentnoniid}. \solution \ still performs the best among the five algorithms, and the other four algorithm are completely defeated by this attack. \solution \ performs worse than it under IID scenario due to the data distribution. Even so, \solution \ is still effective against this attack.

\begin{figure}[!h]
	\centering
	\includegraphics[width=3in]{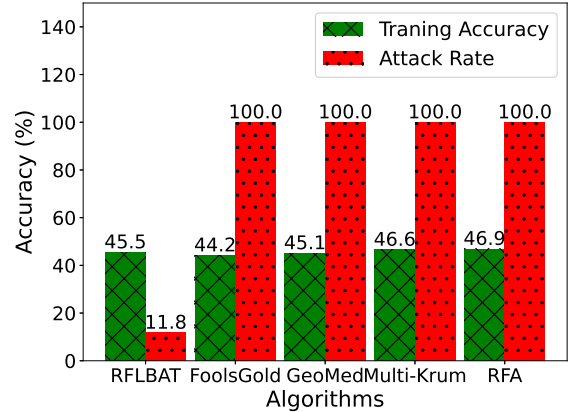}
	\caption{The performance of FoolsGold, Multi-Krum, GeoMed, RFA algorithms and \solution \ on Amazon under Non-IID scenario with 90\% backdoored clients.}
	\label{sentimentnoniid}
\end{figure}

\subsubsection{The Experiment Results of DNS}
\solution \ relies on the the key sight that the backdoored clients perform more similar than benign clients. However, similarity is greatly influenced by different data distributions. To test the effectiveness of \solution \ under diverse data distribution, we implement a DNS attack by changing Dirichlet distribution on MNIST.

Figure \ref{data_distribution} shows the performance of \solution \ against DNS attack. We see that  \solution \ can effectively defend against this attack with low attack rate and satisfied training accuracy. It indicates that \solution \ can distinguish backdoored gradients and benign gradients and select benign gradients to aggregate under various Non-IID scenarios. Note that \solution \ performs the worst when the Dirichlet distribution parameter $\alpha=0.1$. This is because the data distribution at $\alpha=0.1$ is more heterogeneous than it at $\alpha>0.1$, as shown in Figure \ref{sam}. Specially, from Figure \ref{sam}, we can see that the data distribution of all clients at $\alpha=0.1$ is more diverse than it at $\alpha=2$, resulting in more heterogeneous data distribution at a lower $\alpha$. Due to more heterogeneous data distribution, \solution \ will cluster the gradients into more clusters, and each cluster contains fewer clients. Although \solution \ can effectively select the optimal cluster to aggregate, this cluster only consists of partial clients of all benign clients. Therefore, \solution performs the worst when the Dirichlet distribution parameter $\alpha=0.1$. However, \solution \ performs similar as the Dirichlet distribution parameter $\alpha>0.1$. It indicates that backdoor attack cannot subvert \solution \ by manipulating the data distribution.

\begin{figure}[!h]
	\centering
	\includegraphics[width=3in]{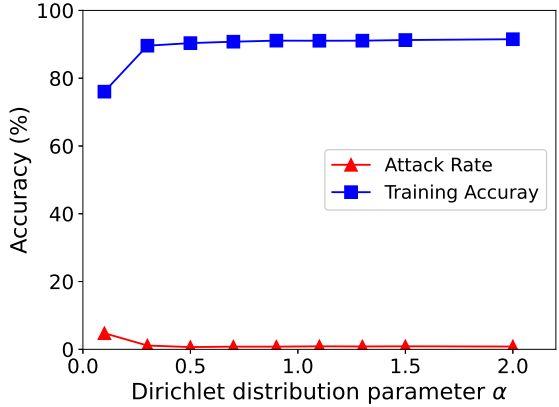}
	\caption{The performance of \solution \ against DNS attack on MNIST.}
	\label{data_distribution}
\end{figure}

\begin{figure}[!h]
	\centering
	\subfigure[$\alpha=0.1$]{
	\label{sam0.1}
	\includegraphics[width=1.6in]{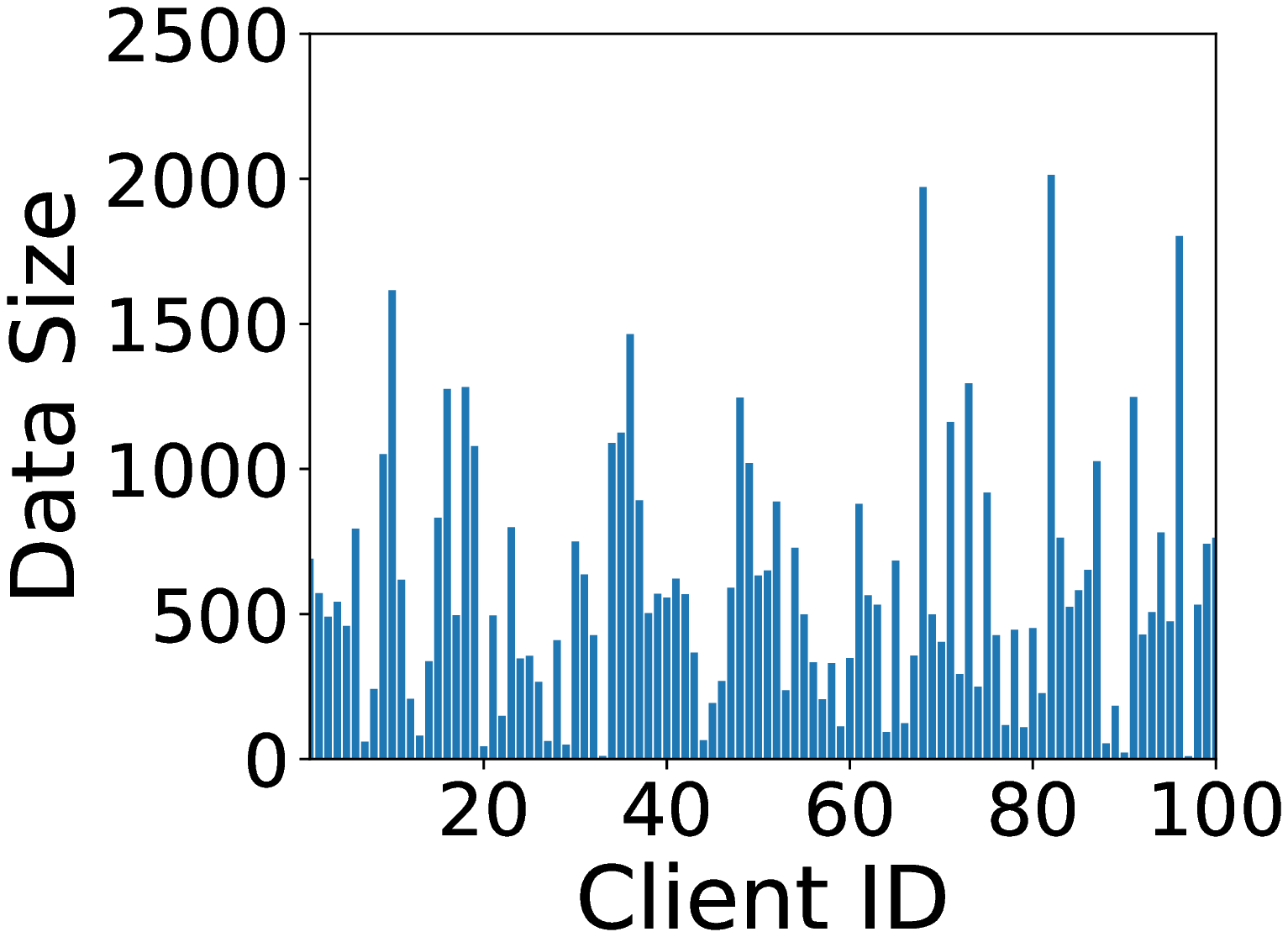}}
	\subfigure[$\alpha=2$]{
	\label{sam2}
	\includegraphics[width=1.6in]{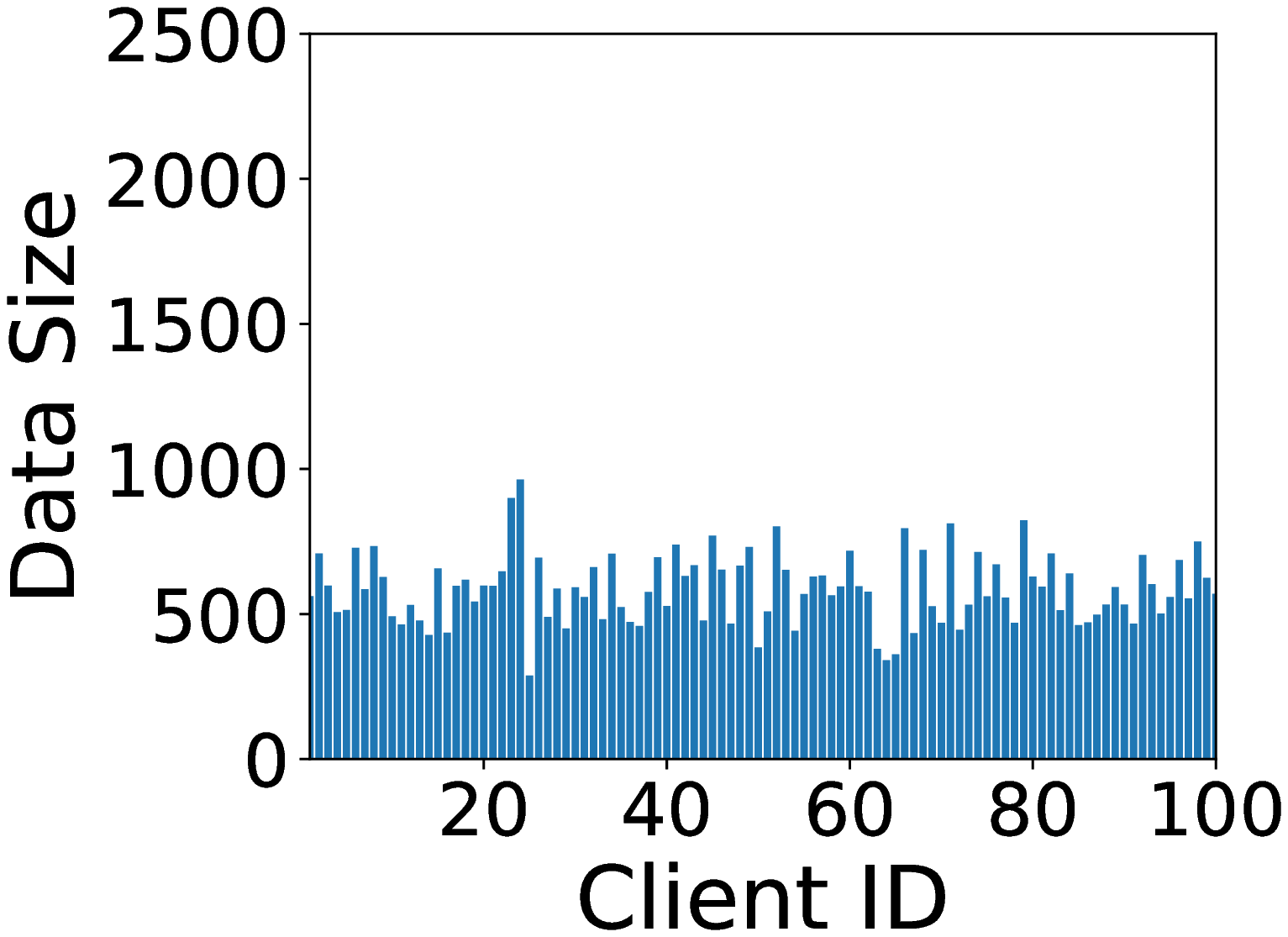}}
	\caption{The data size of all clients under different Dirichlet distributions.}
	\label{sam}
\end{figure}

\subsubsection{The Experiment Results of DNC}
To test the performance of \solution \ facing different number of participating clients, we implement DNC attack with 50\% backdoored clients. We evaluate \solution \ under 50, 100, 200, 400, 600, 800, 1200 and 1600 clients experiments, respectively. Figure \ref{numclients} shows the result of \solution \ under DNC scenario. \solution \ are also effective against DNC attack: the attack rate is only 1\%, while the training accuracy is about 80\% which is not affected by the number of clients. It indicates that the the benign gradients can be clustered into a cluster by PCA technique and Kmeans clustering algorithm and can be selected based on cosine similarity using \solution. Therefore, \solution \  is robust against DNC attack.

\begin{figure}[!h]
	\centering
	\includegraphics[width=3in]{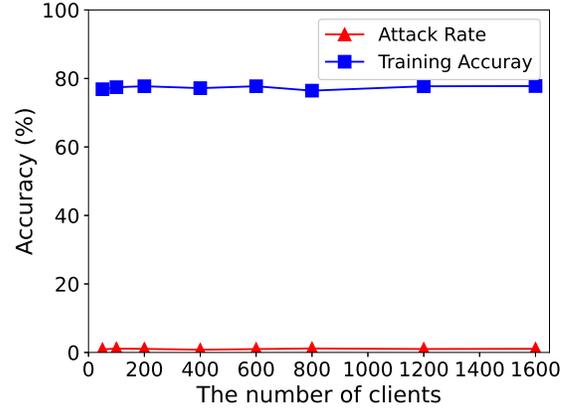}
	\caption{The performance of \solution \ against DNC on FEMNIST.}
	\label{numclients}
\end{figure}

\subsubsection{The Experiment Results of DBA}
Xie \cite{xie2019dba} propose a viewpoint that distributed backdoor attack (DBA) is much stronger than centralized backdoor attack and verify it through a series of experiments. To evaluate the effectiveness of \solution \ under DBA scenario, we also conduct a series of DBA experiments using \solution \ on MNIST, FEMNIST and Amazon datasets. Specially, similar to \cite{xie2019dba}, we separate the centralized backdoor into four parts: backdoor0, backdoor1, backdoor2 and backdoor3, and use each local backdoor to poison 10\% clients among 100 clients. Hence there are 40\% backdoored clients in DBA scenario and each 10\% backdoored clients performs a single backdoor attack. We also consider two scenarios: 1, the backdoored clients push the original gradients to the central server; 2, the backdoored clients amplify the gradients and push them to the central server. Therefore, we set the scale factor $\lambda$ to be 1 and 100 in DBA scenario.

Table \ref{tab:3} shows the training accuracy and attack rate using \solution \ under DBA scenario. We can see that \solution \ can resist DBA, meanwhile maintaining high training accuracy. To be specific, in MNIST, when the scale factor $\lambda=1$, the attack rates of backdoor attacks which contains full backdoor and local backdoor are all about 0.8\%, and the train accuracy is 89.1\%. As the scale factor is 100, the attack rates of backdoor attacks are barely changed, but the training accuracy is 90.2\% which is little higher than that of scale factor $\lambda=1$. The similar results can be found on FEMNIST and Amazon.

\begin{table*}[]
	\caption{The training accuracy and attack rate using \solution \ under DBA scenario on MNIST, FEMNIST and Amazon.}
	\label{tab:3}       
	\begin{center}
		\begin{tabular}{cccccccc}
			\toprule
			
			
			Dataset & Scale factor & Training accuracy & Full backdoor & Backdoor0 & Backdoor1 & Backdoor2 & Backdoor3 \\
			
			\noalign{\smallskip}\hline\noalign{\smallskip}
			
			\multirow{2}*{MNIST} & 1 & 89.1\% & 0.8\% & 0.8\% & 0.8\% & 0.8\% &  0.8\%  \\
								& 100  & 90.2\% & 0.7\% & 0.7\% & 0.6\% & 0.7\% & 0.6\% \\
		
			\multirow{2}*{FEMNIST} & 1 & 66.1\% & 4.1\% & 2.6\% & 2.7\% & 3.4\% & 4.4\% \\
								& 100  & 68.6\% & 2.4\% & 1.6\% & 1.7\% & 1.9\% &  2.3\% \\
						
			\multirow{2}*{Amazon} & 1 & 41.8\% & 5.1\% & 4.5\% & 4.8\% & 4.3\% &  4.6\%  \\
						& 100  & 43.3\% & 3.2\% & 2.8\% & 3.1\% &3.4\% &3.1\% \\
		
			\bottomrule
		\end{tabular}
	\end{center}  	
\end{table*}

To understand how \solution \ handles these two scenarios, we visualize the gradients after PCA for MNIST in Figure \ref{scale}. Figure \ref{scale1} shows the gradients distribution at scale factor $\lambda=1$. Since the gradients are scattered, the benign gradients cannot be clustered into a cluster. \solution \ can only select partial benign gradients to aggregate. In contrast to $\lambda=1$, as the scale factor $\lambda=100$, the benign gradients and backdoored gradients can be clearly distinguished, and the benign gradients can be clustered into a cluster. Hence \solution \ can select all benign gradients to aggregate , resulting in a higher training accuracy.

\begin{figure}[!h]
	\centering
	\subfigure[$\lambda=1$]{
		\label{scale1}
		\includegraphics[width=3in]{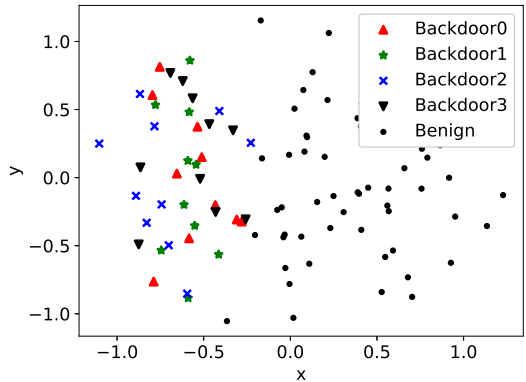}}
	\subfigure[$\lambda=100$]{
		\label{scale100}
		\includegraphics[width=3in]{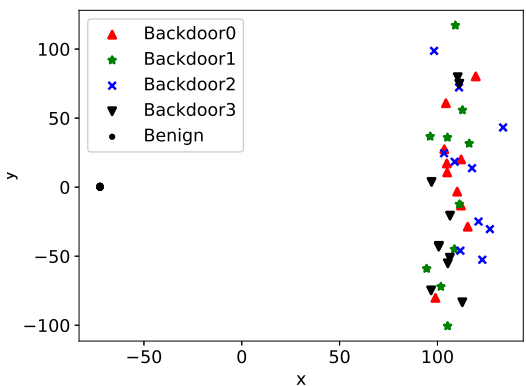}}
	\caption{A visualization of the gradients of clients after PCA under $\lambda=1$ and $\lambda=100$ in MNIST.}
	\label{scale}
\end{figure}

Note that the attack rates of full backdoor and local backdoor are lower at $\lambda=100$ than that at $\lambda=1$. The reason is simple: take MNIST as an example as well, in Figure \ref{scale1} ($\lambda=1$), the backdoored gradients are very close to the benign gradients, leading to a few backdoored gradients in the final selected cluster; in contrast to the result of $\lambda=1$, in Figure \ref{scale100} ($\lambda=100$), the backdoored gradients can be fully detected and removed during aggregation. It is worth mentioning that the multiple benign gradients converge into a point in Figure \ref{scale100}, not a single gradient.

As the results in DBA scenario demonstrate, \solution \ is effective at defending against DBA regardless of whether the backdoored clients amplify the gradients.

\section{Related Works}
\label{sec:Related works}

\textbf{Backdoor Attacks on Federated Learning.} Backdoor attacks aim to insert backdoor into final global model by training strong poisoned local models and submitting model updates to the central server, so as to mislead the final global model \cite{bhagoji2019analyzing}. \cite{bagdasaryan2020backdoor} studies the model replacement approach, where the attacker scales malicious model updates to replace the global model with local backdoored one. \cite{xie2019dba} experimentally proves that distributed backdoor attack is stronger than centralized attack.

\textbf{Robust Aggregation Algorithms in Federated Learning.} To nullify the impact of attacks in aggregating local model updates, many robust aggregation algorithms have been proposed \cite{blanchard2017machine,chen2018distributed,mhamdi2018hidden,pillutla2019robust,yin2018byzantine,li2020learning,2018Mitigating,shen2016auror}. {\color{blue}Krum \cite{blanchard2017machine}} selects a representative worker among multiple workers and use its update to estimate the true update of global model. Bulyan \cite{mhamdi2018hidden} uses Krum to iteratively select benign workers and then aggregates these workers by a variant of the TrimmedMean \cite{yin2018byzantine}. Because the median-based algorithms are more resistant to outliers than mean-based algorithms, other algorithms employ coordinate-wise median \cite{yin2018byzantine}, geometric median \cite{chen2018distributed}, and approximate geometric median \cite{pillutla2019robust} to aggregate a global model. \cite{2019Abnormal,li2020learning} require a pre-trained model to detect and remove malicious model updates in aggregation. The malicious worker detection model can be trained using autoencoder and test data. \cite{shen2016auror} proposes AUROR to address backdoor attacks in collaborative machine learning. \cite{pillutla2019robust} proposes a robust aggregation algorithm named RFA by replacing the weighted arithmetic mean with an approximate geometric median, so as to reduce the impact of the contaminated updates. \cite{2018Mitigating} proposes FoolsGold, which calculates the cosine similarity of the gradient updates from clients, reduces aggregation weights of clients that contribute similar gradient updates, thus promoting contribution diversity.

\section{Conclusion}
In this work, we propose a robust aggregation algorithm named \solution \ based on unsupervised learning against backdoor attack, in which the central server can detect and remove backdoored gradients using PCA technique and Kmeans clustering algorithm. Our algorithm does not require prior knowledge of the expected number of backdoored attackers, and does not access to the training data and test data. We have conducted extensive experiments using MNIST, FEMNIST and Amazon datasets with LR, CNN and LSTM models respectively. We consider four backdoor attack scenarios: different number of attackers (DNA), different Non-IID scenarios (DNS), different number of clients (DNC) and distributed backdoor attack (DBA). The experimental results indicate that \solution \ is able to outperform the existing robust aggregation algorithms, and can mitigate various backdoor attack scenarios. \solution \ is also effective even when backdoored clients overwhelm benign clients.

\bibliographystyle{ijcai21}
\bibliography{ijcai21}
\end{document}